\documentclass[12pt]{iopart}
\usepackage{epsfig}
\usepackage{graphicx}
\newcommand{\bra}{\langle}
\newcommand{\ket}{\rangle}
\newcommand{\minus}{{\!-\!}}
\begin{document}
\def\pin{}
\let\noin=\noindent

\def\Z{{\cal Z}_\beta}
\def\tZ{\tilde{\cal Z}_{\tilde \beta}}
\def\dS{\mbox{\boldmath $S$}}
\def\dT{\mbox{\boldmath $T$}}
\def\ds{\mbox{\boldmath $s$}}
\def\dt{\mbox{\boldmath $t$}}
\def\dm{\mbox{\boldmath $m$}}
\def\ddp{\mbox{\boldmath $p$}}
\def\dq{\mbox{\boldmath $q$}}
\def\bnul{\mbox{\boldmath $0$}}
\def\ptl#1{\frac{\partial}{\partial #1}}
\def\bfm#1{\mbox{\boldmath $#1$}}
\def\IK{{\rm I\!K}}
\def\ID{{\rm I\!D}}
\def\IB{{\rm I\!B}}
\def\IM{{\rm I\!M}}
\def\II{{\rm 1\!\!1}}%{\rm I}}
\def\IE{\mbox{I$\!$E}}
\def\IP{\mbox{I$\!$P}}
\def\IN{\mbox{I$\!$N}}
\def\IR{\mbox{I$\!$R}}

\title[Coupled dynamics in the XY spin-glass]{Coupled dynamics of fast spins and slow exchange
interactions in the XY spin-glass}
\author{
G Jongen \dag, J Anem\"uller \ddag,
D Boll\'e \dag, A C C Coolen \ddag\\
and C P\'erez-Vicente \S}
\address{\dag\ Instituut voor Theoretische Fysica,
            K.U.\ Leuven, B-3001 Leuven, Belgium }
\address{\ddag\
            Department of Mathematics, King's College London,
            The Strand, London WC2R 2LS, UK}
\address{\S Departament de Fisica Fonamental, Facultat de Fisica,
Universitat de Barcelona, 08028 Barcelona, Spain}
\ead{greet.jongen@khleuven.be, joern@anemueller.de,
desire.bolle@fys.kuleuven.ac.be, tcoolen@mth.kcl.ac.uk,
conrad@ffn.ub.es}

\begin{abstract}
We investigate
an XY spin-glass model in which both spins and
interactions (or couplings) evolve in time, but with widely separated
time-scales.
For large times this model can be solved using replica theory,
requiring two levels of replicas, one level for the spins and one
for the couplings. We define the relevant order parameters, and derive a phase
diagram in the replica-symmetric approximation,
which exhibits two distinct spin-glass phases. The first phase is characterized by
freezing of the spins only, whereas in the second phase both spins
and couplings are frozen.
A detailed stability analysis leads also to two
distinct corresponding de~Almeida-Thouless lines, each marking continuous
replica-symmetry breaking.
Numerical simulations support our theoretical study.
\end{abstract}

\pacs{75.10.Nr, 05.20.-y, 64.60.Cn}
%{\bf Key words:} spin glasses, XY model, AT-line, double replica

\section{Introduction}

The study of coupled dynamics of fast Ising spins and slow couplings has
received considerable interest recently (see e.g. \cite{CPS}-\cite{C} and
references therein), stimulated by considerations of
simultaneous learning and retrieval in recurrent neural networks and the
influence of slow atomic diffusion processes in disordered magnetic
systems.

Generalizing spin systems by taking their interactions to be (slowly) time
dependent was first considered in \cite{Ho84}, as a mechanism with which to restore broken
ergodicity at low temperature in the SK~model \cite{SKb}. Another conceptually similar
process, but now describing slow and deterministic
synaptic modification in neural systems, driven by averages over neuron states, was
first introduced in \cite{Sh}. Explicit stochastic dynamical laws for
the interactions were defined in  \cite{CPS,PS,PCS}, where spin-glass
models with coupled dynamics were studied within replica mean-field theory.
It turned out that the replica dimension in such models  has a direct physical
interpretation as the ratio of two temperatures characterizing the
stochasticity in the spin dynamics and the coupling dynamics, respectively.
Later it was shown that the case of negative replica dimension represents
an over-frustrated system \cite{DFM}. In a similar spirit, neural
network models with a coupled dynamics of fast neurons and slow neuronal connections
were treated in \cite{DH}-\cite{CS}.

In this paper the results previously obtained by others  for Ising spin models are further extended to
a classical XY spin glass with dynamic couplings, whose continuous spin variables
are physically more realistic than Ising ones. Moreover, the XY model is
closely related to models of coupled oscillators \cite{Kuramoto}, of
which the neural network version \cite{FS94} provides a phenomenological
description of neuronal firing synchronization in brain tissue. In
particular, we examine the
effects of including an explicit frozen randomness into the dynamics of
the interaction weights.

The model is solved using the replica formalism. Relevant order
parameters are defined and a phase diagram is obtained upon making the
replica-symmetric Ansatz.
Similarly to the Ising case, we find  two different spin glass phases
in addition to a paramagnetic phase.
One spin-glass phase exhibits freezing of the spins in random
directions,  but on the time-scale of the coupling dynamics these `frozen
directions' still continue to change. A second spin-glass phase
exhibits freezing of the spins as well as of the couplings, such that even on the
large time-scales the `frozen directions' of the spins remain stationary.
We perform a detailed stability analysis and calculate
the de~Almeida-Thouless (AT) lines \cite{AT} (of which here there are
two types), where continuous
transitions occur to phases of broken replica symmetry.
A brief preliminary account of the first part of the present work has been
presented in \cite{JBC}. Finally, we
discuss and tackle the problem of simulating this model numerically.

The remainder of this paper is organized as follows. In section 2 the
classical XY spin glass model with coupled dynamics is defined. In
section 3 the order parameters are calculated in the replica
symmetric (RS) Ansatz and a phase diagram is presented. As is well known,
the solutions of this Ansatz are not always stable against replica
symmetry breaking (RSB). Therefore, the lines of instability are calculated in
section 4. Finally, section 5 presents the results of the
numerical simultions of this model, followed by a concluding discussion in
section 6. The appendix describes {\it all} eigenvectors and eigenvalues
of the Hessian matrix determining the stability of the replica symmetric
solutions.

\section{The model}
%\label{secoscill:model}
\pin
We consider a system of $N$ classical two-component spin variables
$\dS_i=(\cos\theta_i,\sin\theta_i)$, $i=1\dots N$, with symmetric
couplings (or exchange interactions) $J_{ij}$, taken to be of infinite
range. In contrast to the standard XY spin glass, these couplings are not
static but are allowed to evolve in time, albeit slowly. The spins are taken to have a
stochastic Glauber-type dynamics such that for {\em stationary} choices
of the couplings the microscopic spin probability density would evolve
towards a Boltzmann distribution
\begin{equation}
   P(\{\dS_i\},\{J_{ij}\}) \sim \exp[-{\beta}{H}(\{\dS_i\},\{J_{ij}\})]
   \label{eq:prob}
\end{equation}
with the standard Hamiltonian
\begin{equation}
H(\{\dS_i\},\{J_{ij}\})=-\sum_{k<\ell}J_{k\ell}~\dS_k\cdot\dS_\ell
\label{eq:Hamiltonian}
\end{equation}
and with inverse temperature $\beta=T^{-1}$, where $k,l\in\{1,\ldots,N\}$, and
where, at least for the purpose of the dynamics of the spins,
the $\{J_{ij}\}$ are to be considered as quenched variables.

We remark that this system is equivalent to a system of $N$ coupled
oscillators with phases $\theta_i$ \cite{Kuramoto}, whose time
evolution is described by a Langevin equation
\begin{equation}
        \frac{d}{dt}\theta_i=\sum_j J_{ij} \sin(\theta_j-\theta_i)
                +\sqrt{\frac{2\tau}{\beta}}\xi_i(t) \,,
                       \label{def:evspins}
\end{equation}
where the $\xi_i(t)$ are defined as independent white noise variables, drawn from a Gaussian
probability distribution  with
\begin{equation}
        \left<\xi_i(t)\right>=0 \qquad
        \left<\xi_i(t)\xi_j(t')\right>=\delta_{ij}\delta(t-t')\,.
\end{equation}

In our model, the couplings also evolve in a stochastic manner, partially in
response to the states of the spins and to externally imposed biases.
However, we assume that the spin dynamics is very fast compared to that of the
couplings, such that on the time-scales of the couplings the spins are
effectively in equilibrium (i.e. we take the adiabatic limit).
For the dynamics of the couplings the following Langevin form is
proposed :
\begin{equation}
        \frac{d}{dt}J_{ij}=
                \frac{\left<\dS_i\cdot\dS_j\right>+K_{ij}}{N}
                -\mu J_{ij}
                +\frac{\eta_{ij}(t)}{N^{1/2}}
        \qquad ~~~~~~i<j=1\dots N \, .
        \label{def:evcouplings}
\end{equation}
The term $\left<\dS_i\cdot\dS_j\right>$, representing local spin
correlations associated with the coupling $J_{ij}$, is a thermodynamic
average over the Boltzmann distribution (\ref{eq:prob}) of the spins, given the
instantaneous couplings $\{J_{k\ell}\}$.  No other spins are involved, in
order to retain the local character of the couplings.
We remark that only the thermal averages (or long time averages) of the spin
correlations play a role, rather than the instantaneous
correlations, since the dynamics of the couplings is (by definition) sufficiently  slow.
External biases $K_{ij}\!=\!\mu N B_{ij}$ serve to steer the weights to
some preferred values.
The $B_{ij}$ are chosen to be quenched random variables, drawn
independently {}from a Gaussian probability distribution with mean
$B_0/N$ and variance $\tilde{B}/N$:
\begin{equation}
        p(B_{ij})=\frac1{\sqrt{2\pi\tilde B/N}}\mbox{exp}
        \left[
        -\frac{\left(B_{ij}-B_0/N\right)^2}{2\tilde B/N}
        \right]
        \label{eqoscill:Bij}
\end{equation}
and are thus reminiscent of the couplings in the original SK model \cite{SKb}.
Here, in contrast, the $\{B_{ij}\}$ generate frozen disorder in the dynamics of the couplings.
The decay term $\mu J_{ij}$ in (\ref{def:evcouplings})
is added in order to limit the magnitude of the couplings.
Finally, the terms $\eta_{ij}(t)$ represent Gaussian white noise
contributions,  of zero mean
and covariance $\bra \eta _{ij}(t)\eta _{kl}(t')\ket =
         2{\tilde T}~\delta _{ik}\delta _{jl}\delta (t-t')$,
with associated temperature $\tilde T = {\tilde \beta}^{-1}$.
Appropriate factors of $N$ are introduced in order to ensure non-trivial behaviour in
the thermodynamic limit $N \rightarrow \infty$.

The model exhibits three independent global symmetries, which can
be expressed efficiently in terms of the Pauli spin matrices
$\sigma_x$ and $\sigma_z$:
\begin{equation}
\begin{array}{llll}
{\rm inversion~of~both~spin~axes:}&& \dS_i\to -\dS_i & {\rm
for~all}~i\\
{\rm inversion~of~one~spin~axis:}&& \dS_i\to \sigma_z\dS_i & {\rm
for~all}~i\\
{\rm permutation~of~spin~axes:}&& \dS_i\to \sigma_x\dS_i & {\rm
for~all}~i \, .
\end{array}
\label{eq:symmetries}
\end{equation}
Upon using algebraic relations such as
$\sigma_x\sigma_z\sigma_x=-\sigma_z$
and $\sigma_z\sigma_x\sigma_z=-\sigma_x$ we see that in the
high $T$ (ergodic) regime these three global symmetries generate the
following local identities, respectively:
\begin{equation}
\bra \dS_i\ket=\bnul,~~~~~~~~
\bra \dS_i\cdot\sigma_x\dS_j\ket=0,~~~~~~~~
\bra \dS_i\cdot\sigma_z\dS_j\ket=0 \, .
\label{eq:identities}
\end{equation}

We note that
the stochastic equation (\ref{def:evcouplings}) for the couplings is
conservative, i.e. it can be written as
\begin{equation}
\frac{d}{dt}J_{ij}=-\frac{1}{N}\ptl{J_{ij}} \tilde H(\{J_{ij}\})
+\frac{\eta_{ij}(t)}{N^{1/2}}
\label{eq:conservative}
\end{equation}
with the following effective Hamiltionian for the couplings:
\begin{equation}
        \tilde H(\{J_{ij}\})=-\frac1\beta\log\Z(\{J_{ij}\})
                +\frac12\mu N\sum_{k<\ell}J_{k\ell}^2
                -\mu N\sum_{k<\ell}B_{k\ell}J_{k\ell}\,.
\label{eq:couplinghamiltonian}
\end{equation}
The term $\Z(\{J_{ij}\})={\rm Tr}_{\{\dS_i\}}\exp[\beta
\sum_{k<\ell}J_{k\ell}\dS_k\cdot\dS_\ell]$ in this expression
is the partition function
of the XY spins with instantaneous couplings $\{J_{ij}\}$.
It follows from (\ref{eq:conservative}) that the stationary probability density for
the couplings is also of a Boltzmann form, with the Hamiltonian
(\ref{eq:couplinghamiltonian}),
and that the thermodynamics of the slow system (the couplings) are
generated by the partition function $\tZ=\int \prod_{k<\ell}
dJ_{k\ell}~\exp[-\tilde{\beta}\tilde{H}(\{J_{ij}\})]$, leading to
(modulo irrelevant prefactors):
\begin{equation}
\hspace*{-5mm}
        \tZ=\int \prod_{k<\ell} dJ_{k\ell}
                \left[\Z(\{J_{ij}\}) \right]^n
\exp\left[\mu \tilde\beta N  \sum_{k<\ell}B_{k\ell}J_{k\ell}
-\frac{1}{2}~\mu \tilde \beta N\sum_{k<\ell}J_{k\ell}^2 \right] \, .
       \label{eq:partition function}
\end{equation}
In contrast to the more conventional spin systems with frozen disorder, where the
replica dimension $n$ is a dummy variable, here we find that $n$ is given by the ratio
$n=\tilde\beta/\beta$, and can
take any real non-negative value. The limit
$n\rightarrow0$ corresponds to a situation in which the coupling
dynamics is driven purely by the Gaussian white noise, rather than by
the spin correlations. Therefore, in this limit the model is equivalent
to the XY model with stationary couplings formulated, as in \cite{KS}.
For $n=1$ the two characteristic temperatures
are the same, and the theory reduces to that corresponding to the exchange interactions being
annealed variables. In the limit $n\rightarrow\infty$ the influence of
spin correlations on the coupling dynamics dominates, and the couplings $J_{ij}$ only
fluctuate modestly (if at all) around their mean values
$(\langle \dS_i\cdot \dS_j \rangle+K_{ij})/\mu N$.

\section{Statics}
\label{secoscill:statics}

We define the disorder-averaged free energy per site
\begin{equation}
 \tilde f=-\frac{1}{\tilde{\beta} N}\bra \log\tZ\ket_B,
\end{equation}
in which
$\bra \cdot \ket_B$ denotes an average over the $\{B_{ij}\}$.
We carry out this average using the identity
$\log \tZ=\lim_{r\rightarrow0}r^{-1}[\tZ^r\!-1]$, evaluating the
latter by analytic continuation {}from integer $r$. Our system, characterized by the
partition function $\tZ$, is thus replicated $r$ times;
we label each replica by a Roman index.
Each of the $r$ functions $\tZ$, in turn, is given by
(\ref{eq:partition function}), and involves
$\Z(\{J_{ij}\})^n$ which is  replaced by the product of
$n$ further replicas, labeled by Greek indices.
For non-integer $n$, again analytic continuation is made
from integer $n$.
Therefore, performing the disorder average in $\tilde{f}$ boils down to
performing the disorder average of $[\tZ]^r$, involving
$nr$ coupled replicas of the original system: $\{\dS_i\}\to
\{\dS_{ia}^\alpha\}$, with $\alpha=1\ldots n$ and $a=1\ldots r$. We
obtain
\begin{eqnarray}
&& \hspace{-2.5cm}
        \langle[\tilde{\cal Z}_{\tilde \beta}]^r \rangle_B=
          \int\prod_{i<j}\left\{
          dB_{ij}~p(B_{ij})\right\}
          \int \prod_{i<j}\left\{\prod_{a}dJ_{ij}^a
            \left[ \frac{N}{2\pi\tilde J }\right]^{1/2}\right\}
          \nonumber \\
&&\hspace{-2.cm}\times
        {\rm Tr}_{\{\dS_{ia}^\alpha\}}
        \mbox{exp}\left[
          -\frac{N}{2\tilde J}\sum_{i<j} \sum_a
            (J_{ij}^a)^2
          +\frac{N}{\tilde J}\sum_{i<j}\sum_{a}~ B_{ij}J_{ij}^a
          +\beta\sum_{i<j}\sum_a\sum_{\alpha}
            B_{ij}\dS_{ia}^\alpha \cdot \dS_{jb}^\beta
                  \right]
\label{eq:whole_thing}
\end{eqnarray}
where $\tilde J=1/\mu\tilde\beta$, and with the Gaussian probability distribution
of the external biases $B_{ij}$ as given by eq.
(\ref{eqoscill:Bij}).
The Roman indices ($a,b,\dots$) run {}from $1$ to $r$; the Greek ones
($\alpha,\beta,\dots$) from $1$ to $n$.
Expression (\ref{eq:whole_thing}) can be evaluated  using the standard techniques of replica
mean-field theory \cite{MPV}. Because of the complexity of the replica
structure we indicate the most important steps.
We first perform the integrals over the couplings and the biases,
giving
\begin{eqnarray}
\fl  \langle[\tilde{\cal Z}_{\tilde \beta}]^r \rangle_B=
        {\rm Tr}_{\{\dS_{ia}^\alpha\}}
        \mbox{exp}\left[
          \beta \frac{B_0}{N}\sum_{i<j}\sum_a\sum_\alpha
            \dS_{ia}^{\alpha}\cdot\dS_{ja}^\alpha
          +\beta^2\frac{\tilde B}{N}\sum_{i<j}
            \left(\sum_a\sum_\alpha
              \dS_{ia}^{\alpha}\cdot\dS_{ja}^\alpha\right)^2
        \right.\nonumber\\
          \left.+\frac1{2N}\beta^2\tilde J \sum_{i<j}\sum_a
            \left(\sum_\alpha\dS_{ia}^{\alpha}\cdot\dS_{ja}^\alpha
            \right)^2
                  \right]
\end{eqnarray}
and decouple the $i$- and $j$-components using
\begin{equation}
        \dS_{ia}^\alpha\cdot\dS_{ja}^\alpha ~ ~
          \dS_{ib}^\beta\cdot\dS_{jb}^\beta
        = \frac12 \left(
          (\dS_{ab}^{\alpha\beta})_i\cdot(\dS_{ab}^{\alpha\beta})_j
          +(\dT_{ab}^{\alpha\beta})_i\cdot(\dT_{ab}^{\alpha\beta})_j
        \right) \,.
\end{equation}
Here the quantity $(\dS_{ab}^{\alpha\beta})_i$ is defined as a two-dimensional unit
vector with reference angle equal to the difference of the reference
angles of  $\dS_{ia}^\alpha$ and  $\dS_{ib}^\beta$, whereas
$(\dT_{ab}^{\alpha\beta})_i$ is defined as a two-dimensional unit vector with reference
angle equal to the sum of both these angles.
Upon applying the saddle-point method in the thermodynamic limit
$N\rightarrow \infty$ we then arrive at
\begin{eqnarray}
 \fl
        \langle\left[\tilde{\cal Z}_{\tilde \beta}\right]^r \rangle_B=
          \mbox{exp}\left[N~\mbox{extr}~
            F(\{\dm_a^\alpha\},\{\ds_{a}^{\alpha\beta}\},
              \{\ds_{ab}^{\alpha\beta}\},\{\dt_{a}^{\alpha\beta}\},
              \{\dt_{ab}^{\alpha\beta}\})\right]
              \label{eq:addition} \\
\fl
        F(\{\dm_a^\alpha\},\{\ds_{a}^{\alpha\beta}\},
          \{\ds_{ab}^{\alpha\beta}\},
          \{\dt_{a}^{\alpha\beta}\},\{\dt_{ab}^{\alpha\beta}\})=
          -\frac18\tilde B\beta^2\sum_{a\neq b}\sum_{\alpha\beta}
            \left((\ds_{ab}^{\alpha\beta})^2
                 +(\dt_{ab}^{\alpha\beta})^2 \right)
          \nonumber \\
          -\frac18\beta^2(\tilde B + \tilde J)
            \sum_a\sum_{\alpha\neq\beta}
            \left((\ds_{a}^{\alpha\beta})^2
                 +(\dt_{a}^{\alpha\beta})^2 \right)
          \nonumber \\
          -\frac12\beta B_0\sum_a\sum_\alpha(\ds_\alpha^a)^2
          -\frac12\beta^2 \tilde J\sum_a\sum_\alpha(\dt_\alpha^a)^2
          \nonumber \\
          +\log G(\{\dm_a^\alpha\},\{\ds_{a}^{\alpha\beta}\},
          \{\ds_{ab}^{\alpha\beta}\},\{\dt_{a}^{\alpha\beta}\},
          \{\dt_{ab}^{\alpha\beta}\})
        \label{eq:free-energy-functional} \\
\fl
        G(\{\dm_a^\alpha\},\{\ds_{a}^{\alpha\beta}\},
          \{\ds_{ab}^{\alpha\beta}\},\{\dt_{a}^{\alpha\beta}\},
          \{\dt_{ab}^{\alpha\beta}\})=
        {\rm Tr}_{\{\dS_a^\alpha\}}~\mbox{exp}\left[
          \beta B_0\sum_a\sum_\alpha
            \dm_a^\alpha\cdot\dS_a^\alpha
        \right.\nonumber \\
          +\frac14\tilde B\beta^2\sum_{a\neq b}\sum_{\alpha,\beta}
            \left(\ds_{ab}^{\alpha\beta}\cdot\dS_{ab}^{\alpha\beta}
                 +\dt_{ab}^{\alpha\beta}\cdot\dT_{ab}^{\alpha\beta}
            \right)
        \nonumber \\
          +\frac14(\tilde B+\tilde J)\beta^2
            \sum_a\sum_{\alpha\neq\beta}
            \left(\ds_{a}^{\alpha\beta}\cdot\dS_{a}^{\alpha\beta}
                 +\dt_{a}^{\alpha\beta}\cdot\dT_{a}^{\alpha\beta}
            \right)
        \nonumber \\
          \left.+\frac14\beta^2 \tilde J\sum_a\sum_\alpha
            \dt_a^\alpha\cdot\dT_a^\alpha
         \right]        \,.
\end{eqnarray}
The parameters
$\{\dm_a^\alpha\},~\{\ds_{a}^{\alpha\beta}\},~\{\ds_{ab}^{\alpha\beta}\},
~\{\dt_{a}^{\alpha\beta}\}$ and $\{\dt_{ab}^{\alpha\beta}\}$ introduced by
this procedure are vectors, hence the extremum is taken over both
components. They carry Greek and Roman replica labels.

Those parameters which have only one Greek and Roman replica label
($\dm_{a}^{\alpha},~\dt_{a}^{\alpha}$), can be interpreted as
\begin{equation}
        \bfm{m}_a^\alpha
          =\lim_{N\to\infty}\frac1N\sum_{i}
          \left<~\overline{\left<
          \dS_{ia}^{\alpha}\right>}~\right>_{\!B} \qquad
        \bfm{t}_a^\alpha
          =\lim_{N\to\infty}\frac1N\sum_{i}
          \left<~\overline{\left<
          \dT_{ia}^{\alpha}\right>}~\right>_{\!B} \,.
\end{equation}
The horizontal bar denotes thermal averaging over the
coupling dynamics with fixed biases $\{B_{ij}\}$.
Those parameters which involve pairs of replicas either connect two distinct Greek
replicas  with a single Roman replica,
($\ds_{a}^{\alpha\beta},~\dt_{a}^{\alpha\beta}$), or with two distinct Roman
replicas, ($\ds_{ab}^{\alpha\beta},~\dt_{ab}^{\alpha\beta}$).
The latter vector variables can, equivalently, be expressed in terms of
 the following scalar order parameters,
which measure the correlations between the various replicas:
\begin{eqnarray}
        q_{ab}^{\alpha\beta}&=&\lim_{N\to\infty}\frac1N\sum_{i}
          \left<~\overline{\left<
          \dS_{ia}^{\alpha}\cdot\dS_{ib}^{\beta}\right>}~\right>_{\!B}
\nonumber \\
        u_{ab}^{\alpha\beta}&=&\lim_{N\to\infty}\frac1N\sum_{i}\left<~\overline{\left<
          \dS_{ia}^{\alpha}\cdot
          \sigma_x \dS_{ib}^{\beta}
          \right>}~\right>_{\!B}
\nonumber \\
        v_{ab}^{\alpha\beta}&=&\lim_{N\to\infty}\frac1N\sum_{i}\left<~\overline{\left<
          \dS_{ia}^{\alpha}\cdot
          \sigma_z \dS_{ib}^{\beta}
          \right>}~\right>_{\!B}  \label{eq:scal.order-parameters}\, .
\end{eqnarray}
At this point we remark that the order parameters $u_{ab}^{\alpha\beta}$ and
$v_{ab}^{\alpha\beta}$ are typical for the XY-model \cite{KS}, and do not
appear in the SK-model;
comparison with (\ref{eq:identities}) shows that, together with
$\bfm{m}_a^\alpha$ and $\bfm{t}_a^\alpha$, they
measure the breaking of the global symmetries (\ref{eq:symmetries}).
For simplicity we will henceforth choose $B_0=0$. We will make the usual
assumption that, in the absence of global symmetry-breaking forces,
phase transitions can lead to at most {\em local} violation of the
identities (\ref{eq:identities}). Thus the latter will remain valid
if averaged over all sites, at any temperature, which implies
that $\bfm{m}_a^\alpha =\bfm{t}_a^\alpha \!=\! \bnul$ and
that $u_{ab}^{\alpha\beta}\!=\!v_{ab}^{\alpha\beta}\!=\!0$.
The spin-glass order parameters $q_{ab}^{\alpha\beta}$,
on the other hand, are not related to simple global symmetries, but
measure the overlap of two vector spins, and serve to characterize the
various phases.

At this stage in the calculation we make the replica symmetry (RS) Ansatz.
Since observables with identical Roman indices refer to system copies with
identical couplings, whereas observables with identical Roman indices {\em
and} identical Greek indices refer to system copies with
identical couplings {\em and} identical spins, in the present problem the RS ansatz for
 the spin-glass order
parameters takes the form
$q_{ab}^{\alpha\beta}=\delta_{ab}\left\{\delta_{\alpha\beta}+q_1[1-
\delta_{\alpha\beta}]\right\}+q_{0}[1-\delta_{ab}]$.
Here we remark that $\dS_{a}^\alpha\cdot\dS_{a}^\alpha=1$ and that, in the
absence of global symmetry breaking forces, $\ds_{ab}^{\alpha\beta}$
becomes a vector of length  $q_{ab}^{\alpha\beta}$ and reference angle $0$.

The asymptotic disorder-averaged free energy per site can now be written as
\begin{eqnarray}
\fl
          \tilde f=\frac18\tilde B \beta^2 n^2 q_0^2
          -\frac18(\tilde B+\tilde J)\beta^2n(n-1)q_1^2
          -\frac14(\tilde B+\tilde J)\beta^2nq_1
        \nonumber \\
          +\int{\cal D}\ddp~\log~\left\{\int{\cal D}\dq~
            \left[{\rm Tr}_{\dS}\mbox{exp}\left[
              \beta~\sqrt{\frac12\tilde B q_0}~\ddp\cdot\dS
                +\beta~\Xi~\dq\cdot\dS
              \right]
            \right]^n
          \right\}      ,
\end{eqnarray}
with the short-hand $\Xi\!=\!
\beta\sqrt{\frac{1}{2}(\tilde{J}\!+\!\tilde{B})q_1\!-
\!\frac{1}{2}\tilde{B}q_0}$,
and where we have introduced the two-dimensional Gaussian measure
\begin{equation}
        {\cal D}\ddp=(2\pi)^{-1}dp_x~dp_y~\mbox{exp}\left[-\frac{1}{2}(p_x^2+p_y^2)\right] \nonumber
        .
\end{equation}
The remaining two order parameters $q_0$ and $q_1$ are determined as the
solutions of the following coupled saddle-point equations
\begin{equation}
\hspace*{-5mm}
        q_0=\int\!dx~P(x)\left\{
          \frac{\int\!dz~P(z)\left[ I_0(z\Xi)\right]^{n-1}
          I_1(z\Xi)~
          I_1(z x\beta \Xi^{-1}\sqrt{\frac{1}{2}\tilde{B}q_0})}
          {\int\!dz~P(z)\left[ I_0(z\Xi)\right]^{n}
          I_0(z x\beta \Xi^{-1}\sqrt{\frac{1}{2}\tilde{B}q_0})}
          \right\}^2
          \label{eq:saddle1}
\end{equation}
\begin{equation}
\hspace*{-5mm}
        q_1=\int\!dx~P(x)\left\{
          \frac{\int\!dz~P(z)\left[ I_0(z\Xi)\right]^{n-2}
          \left[I_1(z\Xi)\right]^{2}
          I_0(z x\beta \Xi^{-1}\sqrt{\frac{1}{2}\tilde{B}q_0})}
          {\int\!dz~P(z)\left[ I_0(z\Xi)\right]^{n}
          I_0(z x\beta \Xi^{-1}\sqrt{\frac{1}{2}\tilde{B}q_0})}
          \right\}
          \label{eq:saddle2}
\end{equation}
 with $P(x)\!=\!x e^{-\frac{1}{2}x^2}\theta[x]$, and where the
functions $I_n(x)$ are the Modified Bessel functions of integer order
\cite{Abramowitz}.

One can give a simple physical interpretation of these order parameters in
terms of the appropriate averages over the various dynamics
\begin{eqnarray}
        q_{0}&=&
          \lim_{N\to\infty}\frac{1}{N}\sum_i\left<~\overline{\left<\dS_i\right>}^2~\right>_B
        \nonumber \\
        q_1&=&
          \lim_{N\to\infty}\frac{1}{N}\sum_i\left<~\overline{\left<\dS_i\right>^2}~\right>_B
     \, .
\label{eq:meaning}
\end{eqnarray}
It is clear that $0\leq q_{0}\leq q_1\leq 1$.

\begin{figure}[t]
\vspace*{2mm}
%\epsfxsize 6.5cm
%\centerline{
%  \hspace{-1.cm}
%  \rotate[r]{\epsfbox{oscill.eps}}}
\centerline{
\includegraphics[%height=5.5cm,
                width=6.8cm, angle=-90]{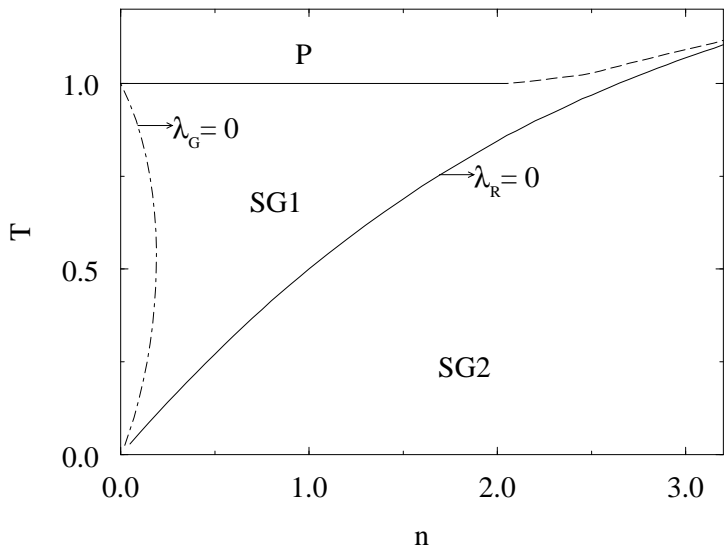}}
\vspace{3mm}
\caption{Phase diagram of the XY~spin~glass with slow dynamic couplings,
drawn in the $n$-$T$ plane with $B_0=0$, $\tilde B=1$ and $\tilde
J=3$. P: paramagnetic phase, $q_1=q_0=0$;
SG1: first spin-glass phase, $q_1>0$ and $q_0=0$ (freezing on spin
time-scales only);
SG2: second spin-glass phase, $q_1>0$ and $q_0>0$ (freezing on all
time-scales);
AT~lines: $\lambda_R=0$ (Roman replicon), $\lambda_G=0$ (Greek replicon).
}
\label{fig:ntxy}
\end{figure}

\begin{figure}[t]
\centerline{
\includegraphics[%height=5.5cm,
                width=5.7cm, angle=-90]{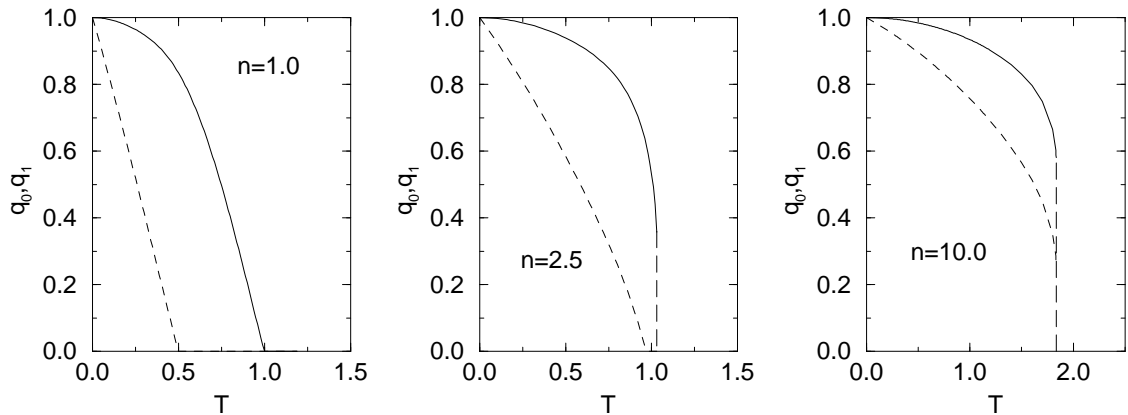}}
%\epsfxsize 5.5cm
%\centerline{\rotate[r]{\epsfbox{orde.eps}}}
\caption{Dependence of the order parameters $q_0$ (broken curve) and $q_1$ (full
curve) on the temperature $T$ for various temperature ratios $n$, all at
$B_0=0$, $\tilde B=1$ and $\tilde J=3$. For $n=1.0$ there is a continuous phase
transition from P to SG1 at $T=0.5$ and from SG1 to SG2 at
$T=1$. For $n=2.5$ the first transition occurs at $T=0.98$ while the
parameter $q_1$ drops discontinuous from $0.18$ to $0$ at $T=1.03$.
Finally for $n=10$ both order
parameters vanish at $T=1.83$ (limit value $q_0=0.13,~q_1=0.29$) indicating a
first order transition from P to SG2. }
\label{fig:ordeparam}
\end{figure}

We have studied the fixed-point equations
(\ref{eq:saddle1},\ref{eq:saddle2}),  after having first eliminated
the parameter redundancy by putting  $\tilde B=1$ and $\tilde J=3$.
The resulting phase diagram in the $n$-$T$ plane is shown in
Fig.~\ref{fig:ntxy}. The appearance of two different spin-glass order
parameters suggests that two different spin-glass phases are to be expected.
Indeed, in addition to a paramagnetic phase (P), where $q_0=q_1=0$,
we find two distinct spin-glass phases: SG1, where $q_{1}>0$ but
$q_0=0$, and  SG2, where both $q_1>0$ and $q_0>0$.
The SG1 phase describes freezing of the spins on the fast time-scales
only (where spin equilibration occurs); on the large time-scales, where
coupling equilibration occurs, we find that, due to the slow motion of
the couplings, the frozen spin directions continually change.
In the SG2 phase, on the other hand, both spins and couplings freeze,
with the net result that even on the large time-scales the frozen spin
directions are `pinned'. The SG1-SG2 transition is always second order and
occurs for $T=(n-1)q_1+1/2$.
The transition SG1-P is second order for
$n<2$ (in which case its location is given by
$\tilde B + \tilde J=4T^2$), but first order for $n>2$.
When $n$ further increases to $n>3.5$, the SG1 phase disappears, and the
system exhibits a first order transition directly from P~to SG2.
Fig.~\ref{fig:ordeparam} shows for several values of $n$
the values of the order parameters as a function of the temperature.

Qualitatively, the phase diagram of the present model is very similar to
that of the Ising spin~glass with dynamic couplings \cite{PS}. The main
difference is the re-scaling by a factor two of the transition temperature
{}from the first spin-glass phase to the paramagnetic phase, as has
already been noticed in \cite{KS}.

The existence of two types of spin-glass order parameters is directly related to the
presence of quenched disorder in the couplings, which allows the latter
to freeze in random directions at low coupling temperature $\tilde T$.
In a model with homogeneous external biases \cite{CPS,Anemuller},
where no preferred direction of the couplings is assumed, one distinguishes
(in contrast to the present situation)
only the paramagnetic phase and the spin-glass phase SG1.
Qualitatively, the transition line separating the paramagnetic phase from  the
spin-glass phase in the case of absent coupling disorder is the same as
that in Fig.~\ref{fig:ntxy}, viz.\ a second
order transition for $n\leq 2$ given by $\tilde J =2T$ and a first order
transition for $n>2$. The corresponding expressions for the order parameters can immediately
be deduced from the results above: when the quenched disorder in the couplings is absent,
the partition function itself is self-averaging and the replica method
is simply no longer needed.
Therefore all order parameters concerning different Roman indices are
redundant and drop out automatically, such that one ends up with
only one spin-glass order parameter. Its explicit value is obtained by
putting  $B_0=0$ and $\tilde B=0$ in (\ref{eq:saddle2}).

\section{Stability of the replica-symmetric solutions}
\label{secoscill:stab}

Additional transitions may occur in our model due to a continuous
breaking of replica symmetry. Here we expect two distinct types of replica symmetry
breaking,
with respect to the two distinct replicas,  viz.\ the Roman and the
Greek ones. The stability of the RS solution is, as always,
expressed in terms of the matrix of second derivatives of quadratic
fluctuations at the saddle point \cite{AT}. We calculate {\it all}
eigenvalues and their multiplicity following the ideas in \cite{PS,AT}.
We remark that our results differ from, and improve upon those of
\cite{PS}. It turns out that the (restricted) set of eigenvectors and eigenvalues given
in \cite{PS} satisfy only part of the relevant orthogonality conditions
used in their calculations. In the following we present a summary of the
results. More details can be found in \ref{App:Hessian_matrix}.

We start by rewriting (\ref{eq:free-energy-functional}), taking into
account its invariance with respect to the global symmetries and the
absence of global symmetry breaking forces
\begin{eqnarray}
\fl     F_S(\{q_{ab}^{\alpha\beta}\},\{q_{a}^{\alpha\beta}\})=
          -\frac18~\tilde B\beta^2\sum_{a\neq b}\sum_{\alpha\beta}
            \left(q_{ab}^{\alpha\beta}\right)^2
          -\frac18~(\tilde B+\tilde J)\beta^2\sum_{a}\sum_{\alpha\neq\beta}
            \left(q_{a}^{\alpha\beta}\right)^2
          \nonumber \\
          +\log~G_S(\{q_{ab}^{\alpha\beta}\},\{q_{a}^{\alpha\beta}\})
        \nonumber \\
\end{eqnarray}
\begin{eqnarray}
\fl
        G_S(\{q_{ab}^{\alpha\beta}\},\{q_{a}^{\alpha\beta}\})=
          {\rm Tr}_{\{\dS_{a}^{\alpha}\}}
          \mbox{exp}
            \left[
              \frac14~\tilde B\beta^2 \sum_{a\neq b}\sum_{\alpha\beta}
                q_{ab}^{\alpha\beta}\dS_a^\alpha\cdot\dS_b^\beta
        \right.\nonumber \\
        \left.\hspace{1.cm}
          +\frac14~(\tilde B+\tilde J)\beta^2\sum_{a}\sum_{\alpha\neq\beta}
                q_{a}^{\alpha\beta}\dS_a^\alpha\cdot\dS_a^\beta
            \right]\,.
        \label{eq:F-global-symmetry}
\end{eqnarray}
We consider small fluctuations of the order parameters around their RS
saddle-point values
\begin{equation}
        q_a^{\alpha\beta}=q_0+\epsilon_a^{\alpha\beta}
        \quad (\alpha<\beta)
        \qquad{\rm and}\qquad
        q_{ab}^{\alpha\beta}=q_1+\eta_{ab}^{\alpha\beta}
        \quad (a<b)
\end{equation}
and expand (\ref{eq:F-global-symmetry}) up to second
order in $\epsilon_a^{\alpha\beta}$ and $\eta_{ab}^{\alpha\beta}$.
The first order terms vanish by construction. The coefficients of the
second order terms form the so-called Hessian matrix and are denoted by
\begin{equation}
     {\cal H}(ab\alpha\beta, cd\gamma\delta)=
                \frac{\partial^2
              F_S(\{q_{ab}^{\alpha\beta}\},\{q_{a}^{\alpha\beta}\})}
                        {\partial q^{\alpha\beta}_{ab}\partial
                                q^{\gamma\delta}_{cd}} ~
                        \rule[-0.4cm]{0.01cm}{1.cm}_{~q_0,q_1}\,.
        \label{eq:hessian_matrix}
\end{equation}
The first argument of ${\cal H}$ (4 components ($ab\alpha\beta$) when
$a\neq b$ and 3 components ($a\alpha\beta$) when $a=b$ but $\alpha\neq\beta$)
denotes the index of the row of the matrix; the last one the column index.
Because of the symmetry of the order parameters
(\ref{eq:scal.order-parameters}) we can always take $a<b$ or
$\alpha<\beta$ when $a=b$. Therefore the square matrix
${\cal H}$ has dimension $\frac12[rn(n-1)+r(r-1)n^2]$.
One can distinguish three groups of matrix elements: firstly,
those related to RSB fluctuations around $q_1$ only,
\begin{eqnarray}
\fl     A_1={\cal H}(a\alpha\beta, a\alpha\beta)
            = -J+J^2
              \left\{
                \left<\overline{\left<\left(\dS_a^\alpha \cdot
                         \dS_a^\beta\right)^2\right>}\right>
                -q_0^2
              \right\}\nonumber\\
\fl     A_2={\cal H}(a\alpha\beta, a\alpha\delta)=
                {\cal H}(a\alpha\beta, a\gamma\beta)
            = J^2
              \left\{
                \left<\overline{\left<\dS_a^\alpha \cdot \dS_a^\beta~
                        \dS_a^\alpha \cdot \dS_a^\delta\right>}\right>
                -q_0^2
              \right\}\nonumber\\
\fl     A_3={\cal H}(a\alpha\beta, a\gamma\delta)
            = J^2
              \left\{
                \left<\overline{\left<\dS_a^\alpha \cdot \dS_a^\beta~
                        \dS_a^\gamma \cdot \dS_a^\delta\right>}\right>
                -q_0^2
              \right\} \nonumber \\
\fl     A_4={\cal H}(a\alpha\beta, c\gamma\delta)
            = J^2
              \left\{
                \left<\overline{\left<\dS_a^\alpha \cdot \dS_a^\beta~
                        \dS_c^\gamma \cdot \dS_c^\delta\right>}\right>
                -q_0^2
              \right\}   \,;
        \label{eq:matricesA}
\end{eqnarray}

\noindent
secondly, those related to fluctuations around $q_0$,
\begin{eqnarray}
\fl     B_1={\cal H}(ab\alpha\beta, ab\alpha\beta)
            = -B+B^2
              \left\{
                \left<\overline{\left<
                        \left(\dS_a^\alpha \cdot \dS_b^\beta\right)^2
                        \right>}\right>
                -q_1^2
              \right\}\nonumber\\
\fl     B_2={\cal H}(ab\alpha\beta, ab\alpha\delta)=
                {\cal H}(ab\alpha\beta, ab\gamma\beta)
            = B^2
              \left\{
                \left<\overline{\left<\dS_a^\alpha \cdot \dS_b^\beta~
                        \dS_a^\alpha \cdot \dS_b^\delta\right>}\right>
                -q_1^2
              \right\}
              \nonumber\\
\fl     B_3={\cal H}(ab\alpha\beta, ab\gamma\delta)
            = B^2
              \left\{
                \left<\overline{\left<\dS_a^\alpha \cdot \dS_b^\beta~
                        \dS_a^\gamma \cdot \dS_b^\delta\right>}\right>
                -q_1^2
              \right\} \nonumber \\
\fl     B_4={\cal H}(ab\alpha\beta, ad\alpha\delta)=
                {\cal H}(ab\alpha\beta, cb\gamma\beta)
            = B^2
              \left\{
                \left<\overline{\left<\dS_a^\alpha \cdot \dS_b^\beta~
                        \dS_a^\alpha \cdot \dS_d^\delta\right>}\right>
                -q_1^2
              \right\}\nonumber\\
\fl     B_5={\cal H}(ab\alpha\beta, ad\gamma\delta)=
                {\cal H}(ab\alpha\beta, cb\gamma\delta)
            = B^2
              \left\{
                \left<\overline{\left<\dS_a^\alpha \cdot \dS_b^\beta~
                        \dS_a^\gamma \cdot \dS_d^\delta\right>}\right>
                -q_1^2
              \right\}\nonumber\\
\fl     B_6={\cal H}(ab\alpha\beta, cd\gamma\delta)
            = B^2
              \left\{
                \left<\overline{\left<\dS_a^\alpha \cdot \dS_b^\beta~
                        \dS_c^\gamma \cdot \dS_d^\delta\right>}\right>
                -q_1^2
              \right\}  \,;
        \label{eq:matricesB}
\end{eqnarray}

\noindent
and finally the matrix elements describing mixed RSB fluctuations
\begin{eqnarray}
\fl     C_1={\cal H}(a\alpha\beta, ad\alpha\delta)
                ={\cal H}(a\alpha\beta,ca\gamma\beta)
            = J~B
              \left\{
                \left<\overline{\left<\dS_a^\alpha \cdot \dS_a^\beta~
                        \dS_a^\alpha \cdot \dS_d^\delta\right>}\right>
                -q_0 q_1
              \right\}\nonumber\\
\fl     C_2={\cal H}(a\alpha\beta, ad\gamma\delta)
                ={\cal H}(a\alpha\beta, ca\gamma\delta)
            = J~B
              \left\{
                \left<\overline{\left<\dS_a^\alpha \cdot \dS_a^\beta~
                        \dS_a^\gamma \cdot \dS_d^\delta\right>}\right>
                -q_0 q_1
              \right\}\nonumber\\
\fl     C_3={\cal H}(a\alpha\beta, cd\gamma\delta)
            = J~B
              \left\{
                \left<\overline{\left<\dS_a^\alpha \cdot \dS_a^\beta~
                        \dS_c^\gamma \cdot \dS_d^\delta\right>}\right>
                -q_0 q_1
              \right\},
        \label{eq:matricesC}
\end{eqnarray}
with
\begin{equation}
   B=\frac14~\tilde B \beta^2 ~~~~~~~~~~~~~~
         J=\frac14~(\tilde B + \tilde J) \beta^2~.
\end{equation}
A simple interpretation of these matrix elements, similar to that  given in
e.g.~\cite{PS}, is not possible here, due to the vector character of the
spins.

\begin{figure}[t]
\vspace*{2mm}
%\epsfxsize 6.5cm
%\centerline{  \hspace{-1.cm}
%\rotate[r]{\epsfbox{oscillinv.eps}}}
\centerline{\includegraphics[%height=0.5\textwidth,
          width=6.5cm,angle=-90]{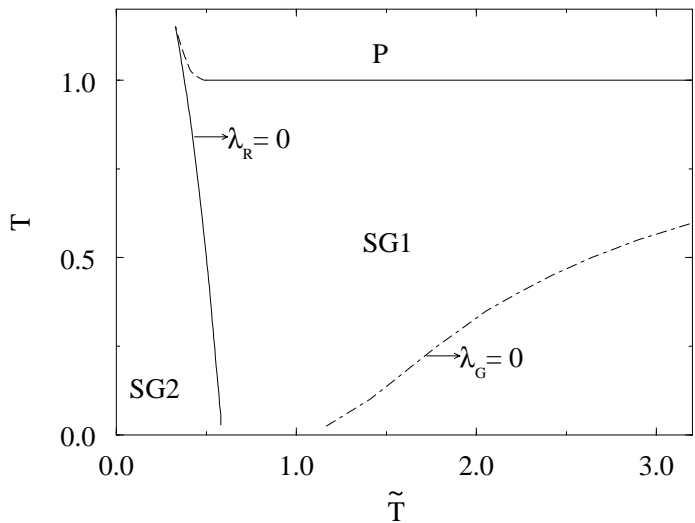} }
\vspace{4mm}
\caption{Phase diagram of the XY~spin~glass with slow dynamic couplings,
drawn in the $\tilde{T}$-$T$ plane; for $B_0=0$, $\tilde B=1$ and $\tilde
J=3$. Further notation as in Fig.~\protect\ref{fig:ntxy}.}
\label{fig:ttildetxy}
\end{figure}

The RS~solutions are stable when the matrix (\ref{eq:hessian_matrix})
is negative definite. Upon analysing all eigenvalues
(see ~\ref{App:Hessian_matrix}) it turns out that only two of
these can cause the occurrence of a region of broken stability. The first replicon eigenvalue,
which we will call the Greek replicon, reads
\begin{equation}
        \lambda_G=A_1-2A_2+A_3
        \label{eq:Greek_eigenvalue}
\end{equation}
and determines the Greek de Almeida-Thouless (AT) line $\lambda_G=0$.
This AT-line measures the breaking of the symmetry with respect to the Greek
indices. The corresponding eigenvectors are given by
eq.~(\ref{eq:eigenvector7}).
The structure of the Greek replicon resembles the one of the replicon
mode of the SK~model found in \cite{AT}, and also the Greek replicon mode of
the SK~model with coupled dynamics as studied in \cite{PS}.
Since the Greek replicas are (by construction) related to the spin
dynamics, the associated region of broken symmetry is located in the region of the phase diagram
with low
temperature $T$.
The Roman replicon eigenvalue is given by
\begin{equation}
\fl     \lambda_R=(B_1-2B_2+B_3)+2n(B_2-B_3-B_4+B_5)+n^2(B_3-2B_5+B_6)
        \label{eq:Roman_eigenvalue}
\end{equation}
and measures the breaking of the symmetry with respect to the Roman replicas.
This occurs at low coupling temperature $\tilde T$.
Similar to the Greek replicon (\ref{eq:Greek_eigenvalue}), the
eigenvectors corresponding to the Roman replicon instability
are symmetric under interchanging all but exactly two -- in
this case Roman -- indices.
It turns out that eigenvalues corresponding to eigenvectors which are symmetric under
interchanging all but a number of indices which is larger than two, whether Roman, Greek or a mixture of both,
can not induce an extra region of broken replica symmetry. Therefore, all regions
where the RS~Ansatz is unstable, are
defined by the locations of the Greek AT~line ($\lambda_G=0$) and the
Roman AT~line ($\lambda_R=0$). These lines are drawn in Fig.~\ref{fig:ntxy}
and Fig.~\ref{fig:ttildetxy}. The latter shows explicitly that there is no re-entrance
{}from the region SG1 to the region with broken replica symmetry (RSB)
when $T$ is varied for fixed $\tilde T$.
The RS~solution is always stable in SG1 with respect to the Roman replicas.
In fact one can show analytically that the Roman AT~line coincides with the
SG1-SG2 transition line.

The RS replica theory developed for our model,
with spin and coupling dynamics on two different time-scales, is reminiscent of that
of the simple XY~model with one step replica symmetry breaking (1RSB). Our
eigenvalues also formally resemble e.g. those describing the stability of the
1RSB solution in the perceptron model \cite{Dorotheyev,WeS}.
Note also that the position of the Roman AT~line in our model is quite different {}from that
in \cite{PS},
although the phase diagrams of both models are qualitatively the same.
The set of eigenvectors given there turn out to satisfy only part of the
required orthogonality relations used in their calculations. An improved
phase  diagram for the SK~model can be found in \cite{JBC}.

Finally we remark that the simpler model with homogeneous biases, mentioned
earlier, does not involve Roman replicas, such that
there appears only a Greek AT~line. The latter line is qualitatively the same as the
one in the model considered here.

\section{Simulations}

\begin{figure}[t]
\vspace*{1cm}
\setlength{\unitlength}{0.6mm}
\begin{picture}(0,137)
\put( 80,10){\epsfxsize=100\unitlength\epsfbox{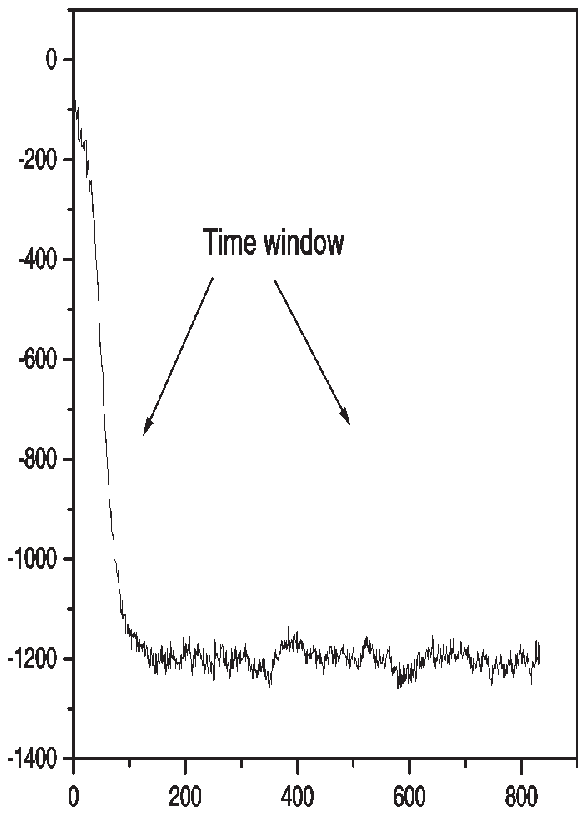}}
\put(133,8){$t$}
\put(75,85){$E$}
\end{picture}
\vspace*{-5mm}
\caption{Evolution in time of the configurational energy (\ref{eq:Hamiltonian}) of the system, for
parameters
$B_0=0$, $\tilde B=1$, $\tilde J=3$, $T=1.1$ and $n=5$, and with a system
of size $N=200$. The
first window is chosen on the basis of the time required
for the combined dynamical system (spins and interactions)
to reach equilibrium; here we decided on a window size of 200.
The values of the observables $q_0$ and $q_1$ were obtained by performing a temporal
average over a (third) time-window of the same size.}
\label{energy}
\end{figure}

%\begin{figure}[t]
%\epsfxsize 6.5cm
%\centerline{  \hspace{-1.cm}
%\rotate[r]{\epsfbox{energy.eps}}}
%\centerline{\includegraphics[%height=0.5\textwidth,
%          width=6.5cm,angle=-90]{energy.eps} }
%\vspace{3mm} \caption{Evolution of the energy of the system. The
%first window gives information about the time required to reach
%equilibrium. To compute the observables $q_0$ and $q_1$ we have
%averaged over the same window size.} \label{energy}
%\end{figure}

\begin{figure}[t]
\vspace*{5mm}
\setlength{\unitlength}{0.6mm}
\begin{picture}(50,137)
\put( 68,17){\epsfxsize=140\unitlength\epsfysize=135\unitlength\epsfbox{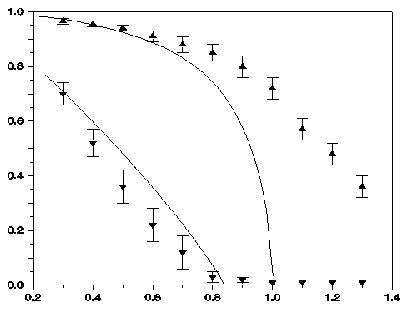}}
\put(137,8){$T$}
\put(65,83){$q_0$, $q_1$}
\end{picture}
\vspace*{-5mm}
\caption{Spin-glass order parameters $q_0$ (circles) and $q_1$ (squares) versus
temperature, for  $n=2$. Continuous lines
represent the theoretical predictions, while symbols denote
simulation results (with $N=200$, averaged over the time-window indicated
in figure \ref{energy} and over 10 samples).
As in the previous figures: $B_0=0$, $\tilde B=1$ and $\tilde J=3$. }
\label{observables_n2}
\end{figure}

\begin{figure}[t]
\vspace*{8mm}
\setlength{\unitlength}{0.6mm}
\begin{picture}(50,137)
\put( 85,10){\epsfxsize=105\unitlength\epsfbox{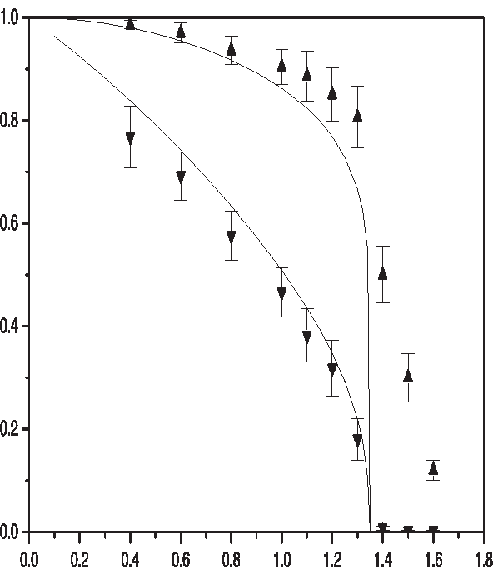}}
\put(137,8){$T$}
\put(65,83){$q_0$, $q_1$}
\end{picture}
\vspace*{-5mm}
\caption{Spin-glass order parameters $q_0$ (circles) and $q_1$ (squares) versus
temperature, for $n=5$. Continuous lines
represent the theoretical predictions, while symbols denote
simulation results (with $N=200$, averaged over the time-window indicated
in figure \ref{energy} and over 10 samples).
As in the previous figures: $B_0=0$, $\tilde B=1$ and $\tilde J=3$. }
\label{observables}
\end{figure}

\begin{figure}[t]
\vspace*{1cm}
\setlength{\unitlength}{0.6mm}
\begin{picture}(0,137)
\put( 80,10){\epsfxsize=100\unitlength\epsfbox{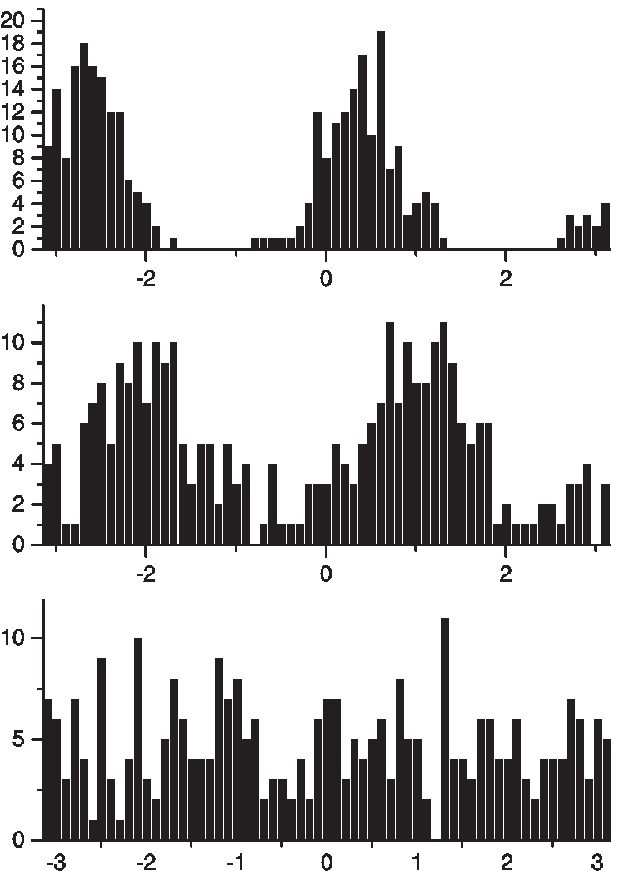}}
\put(132,0){$\phi$}
\put(59,82){$P(\phi)$}
\end{picture}
\caption{Non-normalised phase distribution $P(\phi)=\sum_i\delta[\phi-\phi_i]$
(discretised to a histogram) as observed  at three
different stages of the dynamical process towards equilibrium,
for parameters $B_0=0$, $\tilde B=1$, $\tilde J=3$, $T=1.1$ and $n=5$, and with a system
of size $N=200$.
Bottom graph: (random) phase distribution at $t=0$.
Middle graph: phase distribution at $t=100$ (during transient stage). Top graph: phase distribution
at $t=200$.}
\label{phase}
\end{figure}

In  order to complete our study and verify the predictions of our theory,
we have performed numerical simulations of
our model  (note that, due to the parameter redundancy in our model, we can always restrict ourselves to
$\tilde{J}=3$ and $\tilde{B}=1$). We have considered a population of XY spins, evolving
according to the coupled Langevin equations
(\ref{def:evspins}) and (\ref{def:evcouplings}), which were
discretised according to a standard Euler method, with iteration time step $\Delta t=0.001$.
Several interesting and subtle aspects arise when one attempts to
carry out numerical  simulations of models of the type studied here, with
its widely disparate time-scales.
Firstly, it will be clear that the presence of
 two adiabatically separated time-scales induce extremely large computing
times, which prevent us from numerical exploration of the equilibrium
regime for large system sizes.
This is a general and systematic constraint, which
causes important finite size effects, mainly near the phase transitions.
In all our numerical studies we have, as a result, been forced to
restrict ourselves to relatively modest systems of $N=200$ spins.
A second
point concerns the evolution of the relevant quantities of the
problem, spins and couplings, in view of the need to  calculate
the two main observables of the problem through an averaging
process. This will have to be done very carefully, in order for the measured objects
to indeed be identical to (or at least an acceptable approximation of) those
calculated in the theory. Again the problem is related to having a
finite system size: this narrows the time window where, on the one hand, the fast
processes can be assumed to have been equilibrated, yet, on the other hand,
 the slow processes can be assumed not to have taken place.
Thirdly, there is the fundamental problem that in regions where
replica symmetry no longer holds (beyond either of the two AT
lines) already the spin dynamics will exhibit the traditional
phenomena associated with ageing, including extremely slow
relaxation towards equilibrium; even without the additional
superimposed slow dynamics of the couplings, it would have been
extremely difficult to carry out numerical simulations that would probe
the true equilibrium regime.

We have dealt with these practical problems by adopting the
following strategy. For a given set of couplings we first let spins to relax
to their stationary state; then we perform the average
$\bra\dS_i\cdot\dS_j\ket$ over a number of
time steps sufficiently large to have a statistically reliable  measurement.
We subsequently modify the interactions $\{J_{ij}\}$ for
a certain number of time steps, completing what we call a 'dual
iteration step'. This dual process is repeated until the global
equilibrium state is reached. The key questions in the adequate
employment of this strategy is to quantify rationally the various durations.
 According to the theory, since the time-scale
associated with the couplings is infinitely slow compared to that of the fast
variables (the spins), in each dual updating step we should
modify the interactions $\{J_{ij}\}$ only very slightly (and during only a small number
of update steps $\Delta t$). However, there are limits in practice to the
extent to which one can proceed in this manner, in view of the
danger of the simulations becoming so slow that
they exceed by far one's computing resources. In our simulations
we have updated the interactions during several hundred steps $\Delta t$
(after having satisfied ourselves experimentally that
a duration somewhere between 500 and 1000 iteration steps
is quite appropriate) before, in turn, allowing the spin states  to
evolve. In this manner we have managed to speed up the convergence process
towards global equilibrium, whilst continually verifying
that the stationary values of the order parameters $q_0$ and $q_1$, thus obtained, are not significantly affected.
Even more delicate is deciding on the amount of time during which  to
evaluate the spin averages occurring in  the stochastic equations for the interactions.
If the number of time steps used to calculate these averages is
too large, the spins will have enough time to diffuse over the whole
circle (due to finite size fluctuations which would have been absent in an
infinitely large system), leading to an underestimation of $q_0$. Our experiments
indicate
that averaging over a period of between 2000 and 3000 iteration steps (of duration $\Delta t$ each) gives reliable
results.
Finally, we have to decide on the window size (the number of dual
updating steps) which we have to average over in order to compute the observables
of the system. The logical approach would appear to be to monitor
the evolution of
quantities such as the energy (\ref{eq:Hamiltonian}), starting from the initial state,
until the
stationary state has been (or at least appears to have been) reached. Figure \ref{energy} shows a
typical numerical experiment. The dynamics towards equilibrium on
this time-scale is ultimately controlled by the slow variables, the couplings. The
spins respond to changes in the couplings in a stochastic master/slave fashion,
and only when the slow variables (the couplings) have reached a stationary stationary state can
we speak about global (thermal) equilibrium. In figure \ref{energy} we see that 200 dual steps suffice
to ensure the absence of the main transient effects.

The observed dependence on temperature of the two spin-glass order parameters, $q_0$ and $q_1$,
 is illustrated in figures \ref{observables_n2} and \ref{observables}, together with the corresponding theoretical predictions.
 We have carried out these numerical simulations for the temperature ratios $n=T/\tilde{T}=2$ and $5$,
 respectively.
For the smaller
temperature ratio $n=2$ we observe that our simulations indeed confirm the
existence of two spin-glass phases; one exhibiting freezing
only on spin time-scales, and a second spin-glass phase where
freezing is observed on all time scales. However, quantitative
agreement between theory and simulations is extremely difficult to achieve, due to the practical
problems outlined above.
For $n=5$ the system
is in the region where the theory predicts that a first order phase transition from a
paramagnetic phase to a second-spin glass phase should be found. We
observe a good agreement between theory and simulations, except
for temperatures close to the transition, where finite size effects are obviously
increasingly important.
In addition to the above equilibrium observables, we
have investigated other, non-equilibrium, aspects of our model, by way of further
illustration. We
have measured, for instance, the distribution
$P(\phi)=\sum_i\delta[\phi-\phi_i]$
of phases $\phi_i$, defined via $\dS_i=(\cos\theta_i,\sin\theta_i)$.
 Figure \ref{phase} shows this distribution at three different
 stages during the evolution towards equilibrium, for
 system parameters identical to those of figure \ref{observables}.
Initially the phases were distributed uniformly; one observes this distribution
to deform spontaneously into a bi-modal one, driven in conjunction with the feedback
provided by the (slow) dynamics of the couplings.

\section{Concluding Discussion}
\label{secoscill:conclusions}

\pin
In this paper we have discussed and solved a version of the
classical XY~spin-glass model in which both the spins and their
couplings evolve stochastically, according to coupled equations,
but on widely disparate  time-scales.
The spins play the role of fast variables, whereas the couplings
evolve only very slowly, but according to local stochastic laws which involve the
states of the spins. In the context of disordered magnetic systems this
model describes a situation where one takes into account the possible effects of
slow diffusion of the magnetic impurities, without necessarily assuming energy
equi-partitioning between the slow variables (the impurity locations) and the spins
(hence the potentially different temperatures associated with
each). Alternatively, in the context of neural  systems this model would describe
coupled neural oscillators \cite{Kuramoto} with autonomous stochastic Hebbian-type synaptic adaptation
on the basis of the degree of firing synchrony of pairs of
neurons.

We have solved our model within the
replica-symmetric (RS) mean-field theory, involving two levels of
replicas: one level related to the (slow) couplings, and one
related to the disorder in the problem (the symmetry breaking
terms in the dynamics of the couplings, representing preferred
random values of the latter).
The solution of our model, in RS ansatz, is mathematically similar
to that of the XY model with static couplings, but with one-step RSB. This
is reminiscent of the general connection between the  breaking of replica
symmetry and the existence of dynamics on many time scales \cite{jort}. We
have discussed in detail the stability of the RS-solutions,
including the calculation of {\it all} eigenvalues and their multiplicities
(details of which can be found in the appendices).
It turns out that two distinct replicon eigenvalues determine the
region of stability,
and thus the region of validity of the RS solution.

The thermodynamic phase diagram is found to exhibit two different spin-glass
phases, one where freezing occurs on all time-scales, and one where freezing occurs
only on the (fast) time-scale of the spin dynamics. We also find
both first- and second-order transitions; the origin of the first order ones is the
positive feedback in the system (compared to a system with stationary spin-couplings)
which is induced by the super-imposed coupling dynamics. As could have been expected,
the physics of the present model  resembles
that of the SK~model with dynamic couplings, apart from  a re-scaling in
temperature and provided an appropriate adjustment of the
calculation of the AT~lines in \cite{PS} is made.
Our calculations show how the methods used for solving the Ising
case can be easily adapted to deal with more complicated spin types, and
in addition illustrates further the robustness of the phase diagrams describing the
behaviour of large spin systems with dynamic couplings.

Numerical simulations present further interesting technical challenges, due
to the existence of adiabatically separated time scales (which requires equilibration
of two different nested stochastic processes in order to test the theory),
in addition to the already highly non-trivial and extremely slow dynamics of the fast (spin) system.
In spite of  the  important finite size effects, which are inevitable given
the practical constraints on
available CPU time, our results show good agreement with the theory
and confirm the main characteristics of the predicted behaviour.

\section*{Acknowledgements}

DB would like to thank the National Fund for Scientific
Research-Flanders (Belgium) for financial support.
ACCC and CPV are grateful for support from the Acciones
Integradas programme (British Council \& Ministerio de
Educaci\'{o}n y Cultura, grant 2235).

\clearpage\appendix
\section{The Hessian matrix}
\label{App:Hessian_matrix}

\subsection{Eigenvectors and eigenvalues}

\pin
In this Appendix we show how to find all eigenvectors, eigenvalues and their
multiplicity of the Hessian matrix (\ref{eq:hessian_matrix}).
We immediately remark that ${\cal H}$ is symmetric, implying that its eigenvectors must be orthogonal, a property
exploited heavily in finding its eigenvalues.

\begin{figure}[t]
%\epsfxsize 8.cm
%\centerline{\rotate[r]{\epsfbox{algem.eps}}}
\centerline{\includegraphics[%height=0.5\textwidth,
              width=8cm,angle=-90]{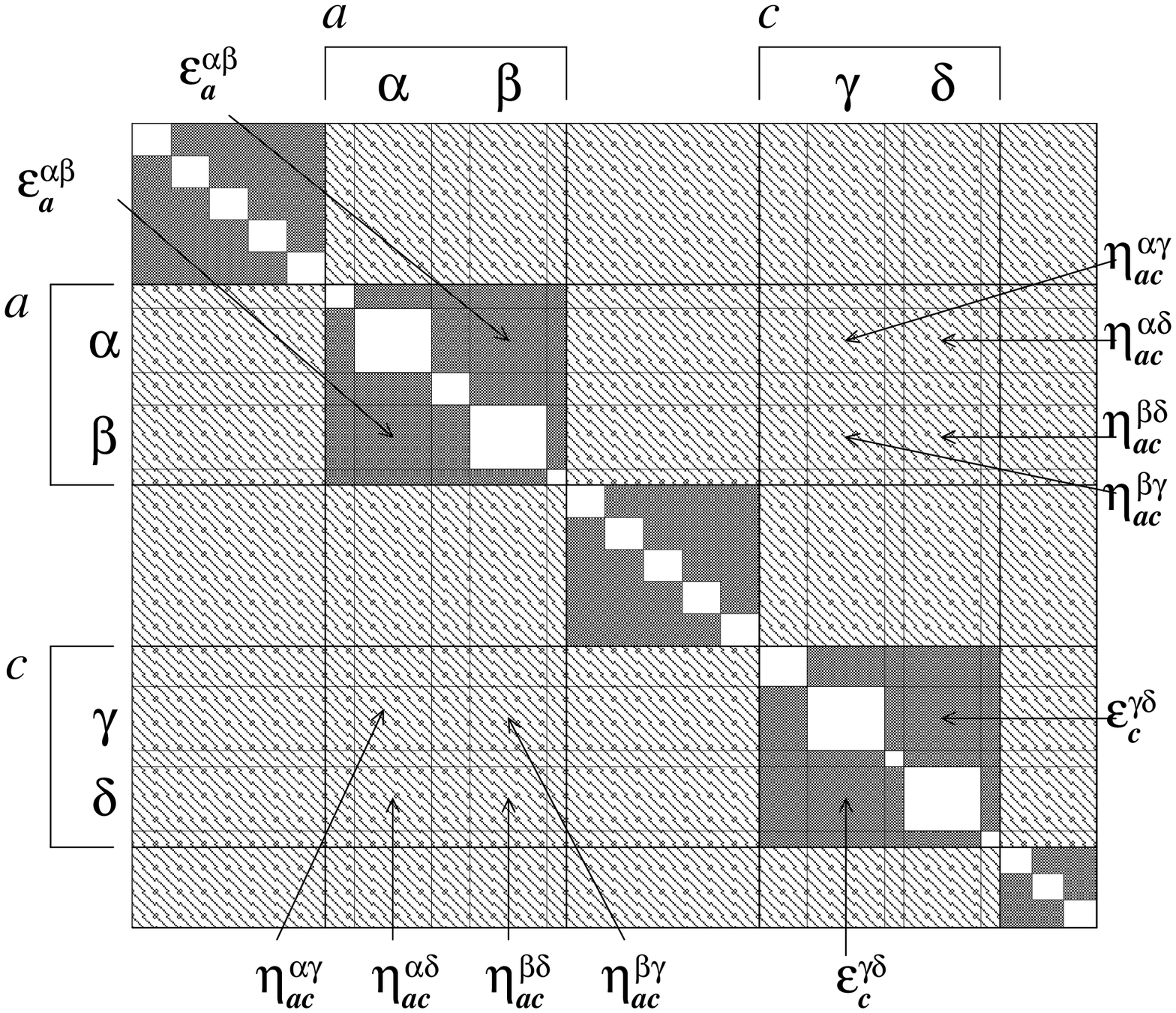}}
\caption{Graphical representation of the structure of a general eigenvector
of the Hessian matrix (\ref{eq:hessian_matrix}). Dark spaces denote the
$\epsilon$-components; the other non-empty spaces are the $\eta$-components.
Further details are found in the text.}
\label{fig:eigenvector_general}
\end{figure}
We denote the eigenvectors in the $\frac12[rn(n-1)+r(r-1)n^2]$-dimensional
space by $(\epsilon_a^{\alpha\beta},\eta_{cd}^{\gamma\delta})$ with
$\alpha<\beta$ and $c<d$, and represent them graphically in a square matrix as
in Fig.~\ref{fig:eigenvector_general}. This matrix is of size $nr\times nr$
and is divided in $r^2$ sub-matrices of size $n\times n$, labeled by two
Roman indices $(a,b,c=1\dots r)$.
The elements of these sub-matrices, in turn, are labeled by two Greek indices
$(\alpha,\beta,\gamma,\delta=1\dots n)$. Thus the rows and columns of the
matrix carry a Roman and a Greek index.
Each of the matrix elements corresponds to a component of the vector
$(\epsilon_a^{\alpha\beta},\eta_{cd}^{\gamma\delta})$, except for the diagonal
components, which are put equal to $0$ (viz.\ an empty space).
The elements of the type $(a\alpha,c\gamma)$ correspond to
$\eta_{ac}^{\alpha\gamma}$ when $a\neq c$ and to $\epsilon_a^{\alpha\gamma}$
when $a=c$ but $\alpha\neq\gamma$. The matrix is symmetric, such that
$\eta_{ac}^{\alpha\gamma}=\eta_{ca}^{\gamma\alpha}$ and
$\epsilon_{a}^{\alpha\gamma}=\epsilon_{a}^{\gamma\alpha}$.

The eigenvectors of ${\cal H}$ with eigenvalue $\lambda$ satisfy
the eigenvalue equation
\begin{equation}
        {\cal H}\left(\begin{array}{c}
                \epsilon_{a}^{\alpha\beta}\\\eta_{cd}^{\gamma\delta}
                \end{array}\right)
        = \lambda \left(\begin{array}{c}
                \epsilon_{a}^{\alpha\beta}\\\eta_{cd}^{\gamma\delta}
                \end{array}\right) \,.
        \label{eq:eigenvalue_equation}
\end{equation}
At this point we remark that $\{\epsilon_a^{\alpha\beta}\}$
are generally uncorrelated fluctuations of the order parameters $\{q_a^{\alpha\beta}\}$ and
$\{\eta_{cd}^{\gamma\delta}\}$ of $\{q_{cd}^{\gamma\delta}\}$. Therefore
we call a vector $(\epsilon_a^{\alpha\beta},\eta_{cd}^{\gamma\delta})$
symmetric under permutation of the Roman and Greek indices when the
components $\{\epsilon_a^{\alpha\beta}\}$ and
$\{\eta_{cd}^{\gamma\delta}\}$ are simultaneously symmetric under
permutation of these indices.
In order to find the explicit form of the eigenvectors we make a general
proposal based on this symmetry. Furthermore, we use the eigenvalue equation
(\ref{eq:eigenvalue_equation}) and the orthogonality of the eigenvectors
corresponding to different eigenvalues.

\begin{figure}[t]
%\epsfxsize 8.cm
%\centerline{\rotate[r]{\epsfbox{iegenv1.eps}}}
\centerline{\includegraphics[%height=0.5\textwidth,
              width=8cm,angle=-90]{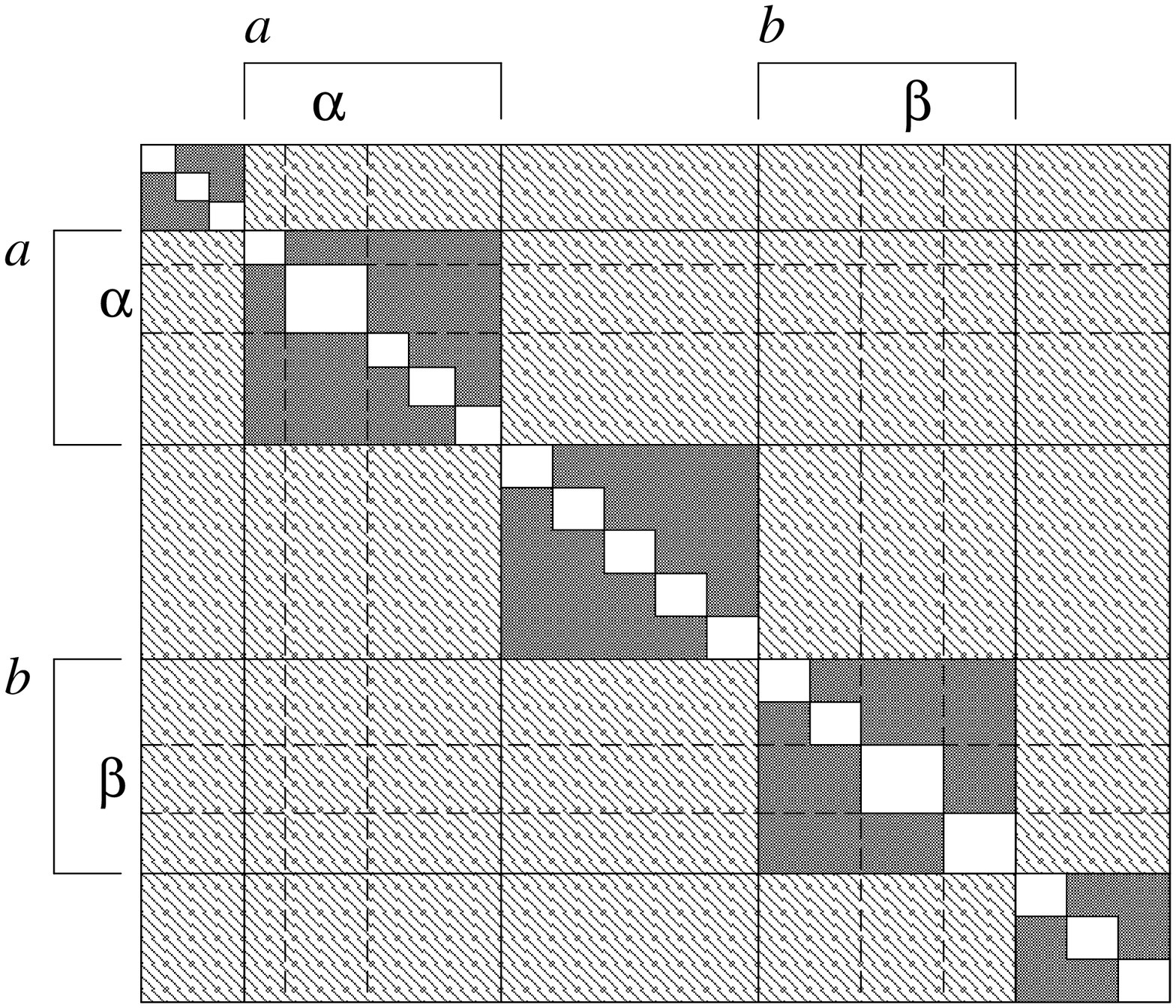}}
\caption{Representation of the replica symmetric eigenvector.
Empty space denotes zero elements; spaces with the same fill pattern denote
elements with identical matrix elements. The dark spaces indicate the
$\epsilon$-components; the other spaces indicate the $\eta$-components. }
\label{fig:rep.symm.eigenvector}
\end{figure}
\begin{figure}[t]
%\epsfxsize 8.cm
%\centerline{\rotate[r]{\epsfbox{iegenv3.eps}}}
\centerline{\includegraphics[%height=0.5\textwidth,
        width=8cm,angle=-90]{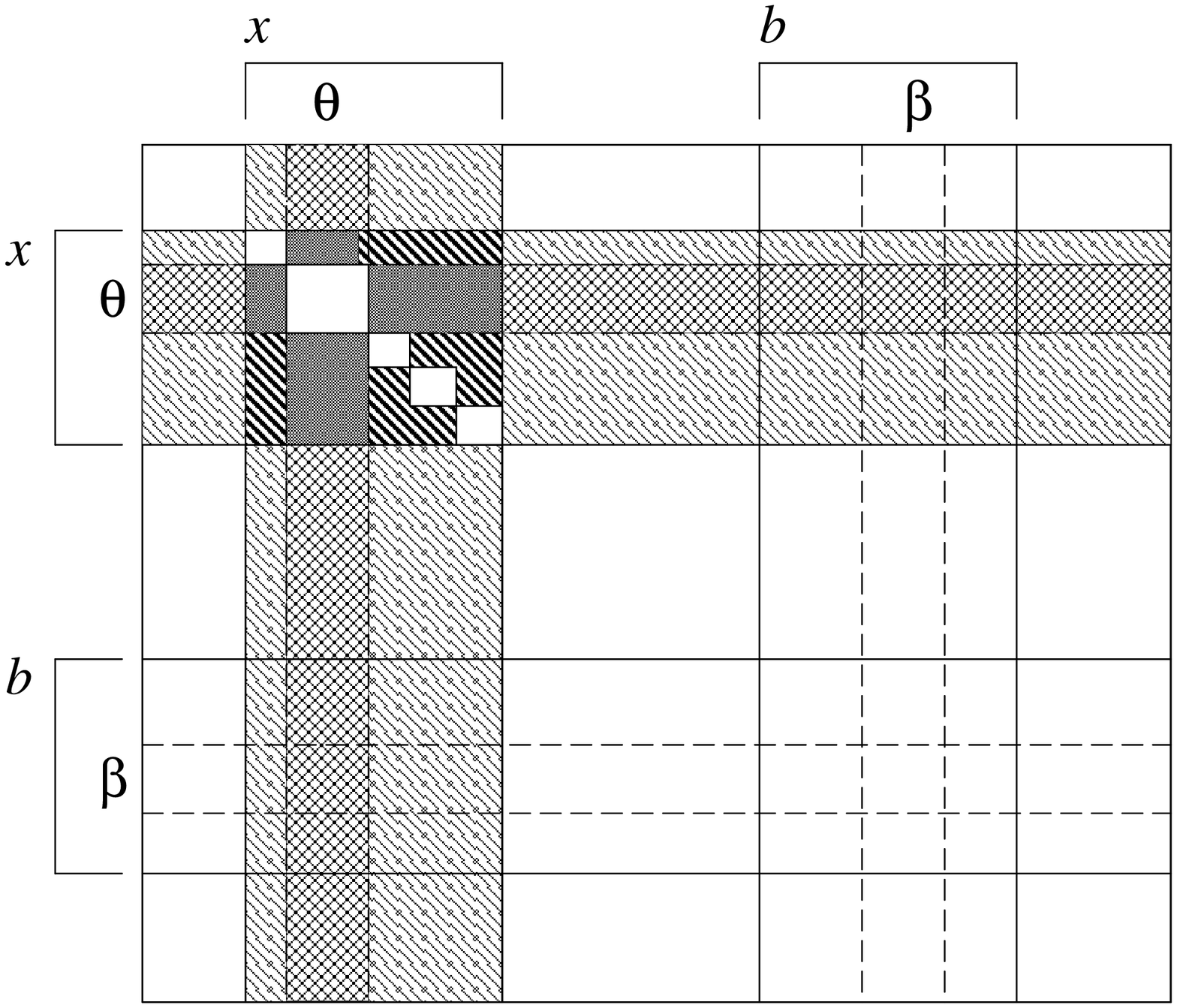}}
\caption{Representation of the eigenvectors (\ref{eq:eigenvector3_4}).
Empty space denotes zero elements; spaces with the same fill pattern denote
elements with identical matrix elements. The dark spaces denote
the components $\epsilon_x^{\theta\beta}$, the striped spaces the
$\epsilon_x^{\alpha\beta}$, the checked  spaces the $\eta_{xb}^{\theta\beta}$,
the other non-empty spaces the $\eta_{xb}^{\alpha\beta}$.}
\label{fig:eigenvector3_4}
\end{figure}

We start from the symmetric solution where all components are identical,
viz.\
\begin{equation}
\epsilon_{a}^{\alpha\beta}=f \quad{\rm and} \quad
                                    \eta_{cd}^{\gamma\delta}=g\,.
\label{eigenvec1}
\end{equation}
These vectors are represented in Fig.~\ref{fig:rep.symm.eigenvector}.
Substitution of (\ref{eigenvec1}) into the eigenvalue
equation (\ref{eq:eigenvalue_equation}) reduces the number of equations to
solve to two and we easily find
\begin{eqnarray}
        &&(X_1-\lambda)~f+Y_1 ~g =0 \nonumber\\
        && X_2 ~f +(Y_2-\lambda)~g=0\nonumber\\
        &&X_1=A_1+2(n-2)A_2+\frac12(n-2)(n-3)A_3
                \nonumber\\
        &&Y_1=2n(r-1)C_1+n(n-2)(r-1)C_2+\frac12n^2(r-1)(r-2)C_3 \,.
\end{eqnarray}
From this we get two non-degenerate eigenvalues
\begin{eqnarray}
        \lambda_{1,2}&=&\frac12\left(Y_2+X_1\pm
                \sqrt{(Y_2+X_1)^2-4(X_1Y_2-X_2Y_1)}\right)
        \label{eq:eigenwaarde1_2}\\
        X_2&=&2(n-1)C_1+(n-1)(n-2)C_2+\frac12n(n-1)(r-2)C_3
                \nonumber\\
        Y_2&=&B_1+2(n-1)B_2+(n-1)^2B_3+2n(r-2)B_4
                \nonumber\\
                &&\hspace{1.cm}
                        +2n(n-1)(r-2)B_5+\frac12n^2(r-2)(r-3)B_6 ~.
                \nonumber
\end{eqnarray}
The matrix elements $A_1,\dots,C_3$
are given by eqs.~(\ref{eq:hessian_matrix},\ref{eq:matricesA}).

The other eigenvalues are related to the breaking of the symmetry in
(\ref{eigenvec1}), both
at the level of Roman indices and at that of the Greek indices. The most simple form of symmetry breaking
is the case
where almost all components are identical, except for those labeled by
a single
specific pair of indices $\{x,\theta\}$:
\begin{eqnarray}
        &&\epsilon_x^{\theta\beta}=f~; \quad
        \epsilon_x^{\alpha\beta}=g~; \quad
        \epsilon_a^{\alpha\beta}=h~; \quad\nonumber\\
        &&\eta_{xb}^{\theta\beta}=k~; \quad
        \eta_{xb}^{\alpha\beta}=l~; \quad
        \eta_{ab}^{\alpha\beta}=m
                \qquad a,b\neq x;~\alpha,\beta\neq\theta
\label{eq:eigenvector3_4alg}
\end{eqnarray}
This increases drastically the number of equations obtained from
(\ref{eq:eigenvalue_equation}).
A first trial solution within this group of candidate eigenvectors is obtained upon
putting $h=m=0$, giving a group of eigenvectors with all components vanishing, except the ones
labeled by $\{x,\theta\}$:
\begin{eqnarray}
        &&\epsilon_x^{\theta\beta}=f~; \quad
        \epsilon_x^{\alpha\beta}=-\frac2{n-2}~f~; \quad
        \epsilon_a^{\alpha\beta}=0~;
        \nonumber \\
        &&\eta_{xb}^{\theta\beta}=k~; \quad
        \eta_{xb}^{\alpha\beta}=-\frac1{n-1}~k~; \quad
        \eta_{ab}^{\alpha\beta}=0
        \nonumber \\
        &&Y_1~k=(\lambda-X1)~f \nonumber \\
        &&X_1=A_1+(n-4)A_2-(n-3)A_3
                \nonumber\\
        &&Y_1=\frac n{n-1}~(n-2)(r-1)\left(C_1-C_2\right)~.
\label{eq:eigenvector3_4}
\end{eqnarray}
The graphical representation of these eigenvectors in the form of a
matrix is drawn in Fig.~\ref{fig:eigenvector3_4}. The associated
eigenvalues read
\begin{eqnarray}
        \lambda_{3,4}&=&\frac12\left(Y_2+X_1\pm
                \sqrt{(Y_2+X_1)^2-4(X_1Y_2-X_2Y_1)}\right)
        \nonumber \\
        X_2&=&(n-1)\left(C_1-C_2\right)
                \nonumber\\
        Y_2&=&B_1+(n-2)B_2-(n-1)B_3+n(r-2)B_4
                \nonumber\\
                &&+n(n-1)(r-2)B_5 \,.
        \label{eq:eigenwaarde3_4}
\end{eqnarray}
The degeneracy of the associated eigenspace is $r(n-1)$; it is found by calculating explicitly
the rank of the matrix composed by these eigenvectors, as will be
outlined in appendix A2.

\begin{figure}[t]
%\epsfxsize 8.cm
%\centerline{\rotate[r]{\epsfbox{iegenv5.eps}}}
\centerline{\includegraphics[%height=0.5\textwidth,
         width=8cm,angle=-90]{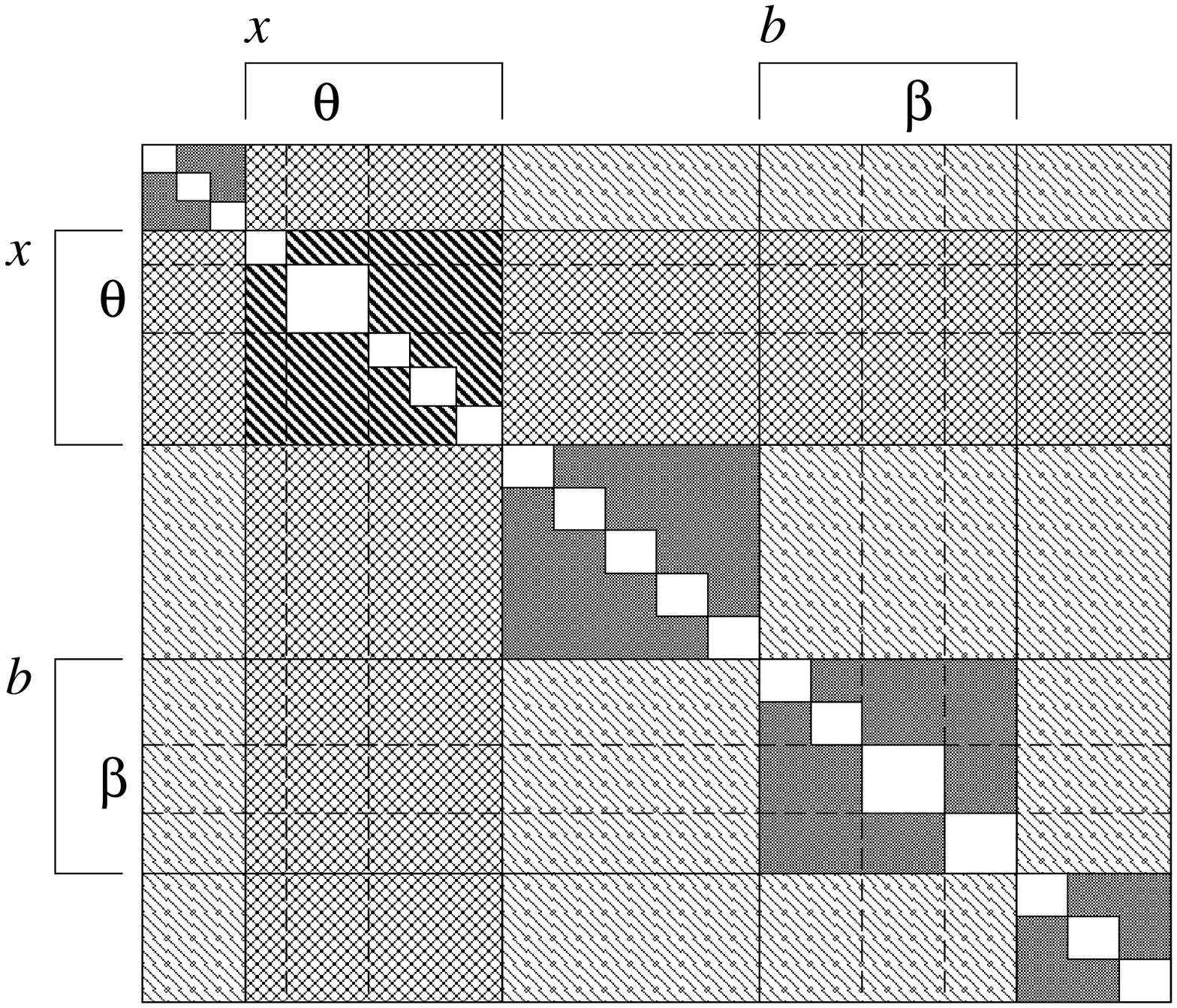}}
\caption{Representation of the eigenvectors (\ref{eq:eigenvectors5_6}).
Empty space denotes zero elements; spaces with the same fill pattern denote
elements with identical matrix elements. Striped spaces denote the
components $\epsilon_x^{\alpha\beta}$, dark spaces the components
$\epsilon_a^{\alpha\beta}$, checked spaces the components
$\eta_{xb}^{\alpha\beta}$, the other non-empty spaces the components
$\eta_{ab}^{\alpha\beta}$.}
\label{fig:eigenvectors5_6}
\end{figure}

Insertion of the general proposal (\ref{eq:eigenvector3_4alg})
into eq. (\ref{eq:eigenvalue_equation}), and subsequently requiring the orthogonality of
this vector to the eigenvectors we have already found earlier, leads us to the new
eigenspace
\begin{eqnarray}
        &&\epsilon_x^{\theta\beta}=
        \epsilon_x^{\alpha\beta}=f~; \quad
        \epsilon_a^{\alpha\beta}=-\frac1{r-1}~f~;
        \nonumber \\
        &&\eta_{xb}^{\theta\beta}=
        \eta_{xb}^{\alpha\beta}=k~; \quad
        \eta_{ab}^{\alpha\beta}=-\frac2{r-2}~k~
        \nonumber \\
        &&Y_1~k=(\lambda-X1)~f\nonumber\\
        &&X_1=A_1+2(n-2)A_2+\frac12~(n-2)(n-3)A_3-\frac12n(n-1)A_4
                \nonumber\\
        &&Y_1=2n(r-1)C_1+n(n-2)(r-2)C_2-n^2(r-1)C_3
        \label{eq:eigenvectors5_6}
\end{eqnarray}
with eigenvalue
\begin{eqnarray}
        \lambda_{5,6}&=&\frac12\left(Y_2+X_1\pm
                \sqrt{(Y_2+X_1)^2-4(X_1Y_2-X_2Y_1)}\right)
        \nonumber\\
        X_2&=&\frac{r-2}{r-1}\left((n-1)C_1+\frac12~(n-1)(n-2)C_2
                -\frac12~n(n-1)~C_3\right)
                \nonumber\\
        Y_2&=&B_1+2(n-1)B_2+(n-1)^2B_3+n(r-4)B_4
        \nonumber\\
        &&%\hspace{1.cm}
                        +n(n-1)(r-4)B_5-n^2(r-3)B_6 \,.
        \label{eq:eigenwaarde5_6}
\end{eqnarray}
The vectors (\ref{eq:eigenvectors5_6}), represented graphically in
Fig.~\ref{fig:eigenvectors5_6}, are symmetric under interchanging  all indices but
one Roman index $x$, and have associated with each eigenvalue a $(r-1)$-dimensional
eigenspace.

\begin{figure}[t]
%\epsfxsize 8.cm
%\centerline{\rotate[r]{\epsfbox{iegenv7.eps}}}
\centerline{\includegraphics[%height=0.5\textwidth,
         width=8cm,angle=-90]{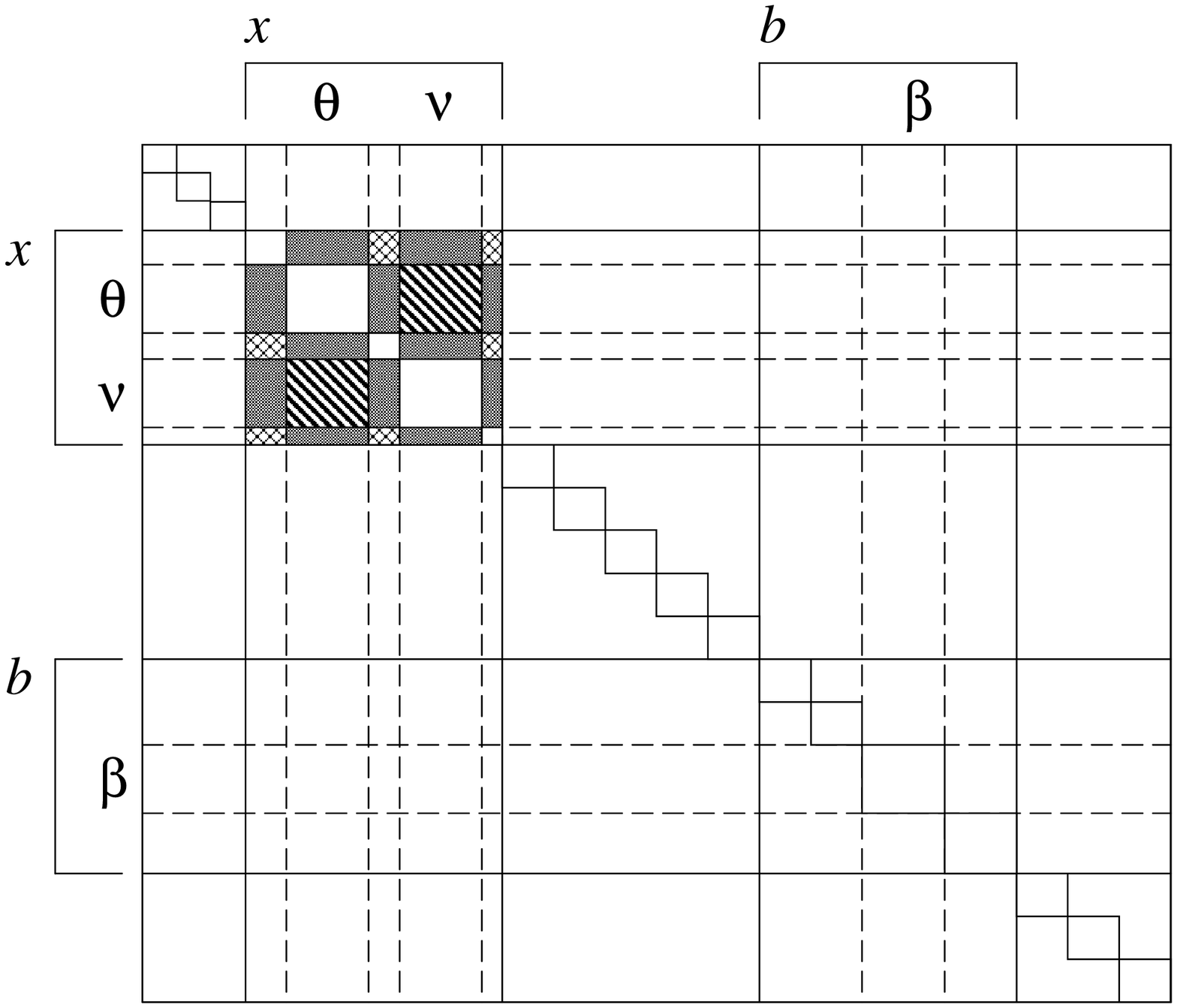}}
\caption{Representation of the eigenvectors (\ref{eq:eigenvector7}).
Empty space denotes zero elements; spaces with the same fill pattern denote
elements with identical matrix elements. the striped spaces indicate the
components $\epsilon_x^{\theta\nu}$, the dark spaces the components
$\epsilon_x^{\theta\alpha}$ and $\epsilon_x^{\alpha\nu}$, the checked spaces
the components $\epsilon_x^{\alpha\beta}$.}
\label{fig:eigenvector7}
\end{figure}
A second class of eigenvectors is found by considering a situation where
one Roman index $x$ and two different indices $\theta$ and
$\nu$ cause breaking of the replica symmetry. In their most general form these are
given by
\begin{eqnarray}
\fl     \epsilon_x^{\theta\nu}=f~; \quad
        \epsilon_x^{\theta\beta}=\epsilon_x^{\alpha\nu}=g~; \quad
        \epsilon_x^{\alpha\beta}=h~; \quad
        \epsilon_a^{\alpha\beta}=k~; \nonumber \\
\fl     \eta_{xb}^{\theta\beta}=\eta_{xb}^{\nu\beta}=l~; \quad
        \eta_{xb}^{\alpha\beta}=m~; \quad
        \eta_{ab}^{\alpha\beta}=p
                \qquad(a,b\neq x;~\alpha,\beta\neq\theta,\nu)~.
\end{eqnarray}
More explicitly, we propose a vector with two special Greek indices, i.e.\ we
try a solution with $k=l=m=p=0$. It corresponds to a vector with all
$\eta$-components and all $\epsilon$-components which are not related to this Roman
index vanishing, and with broken replica symmetry with respect to the two
Greek indices:
\begin{eqnarray}
\fl     \epsilon_x^{\theta\nu}=f~; \quad
     \epsilon_x^{\theta\beta}=\epsilon_x^{\alpha\nu}=-\frac1{n-2}~f~;\quad
     \epsilon_x^{\alpha\beta}=\frac2{(n-2)(n-3)}~f~; \quad
     \epsilon_a^{\alpha\beta}=0; \nonumber \\
\fl     \eta_{xb}^{\theta\beta}=\eta_{xb}^{\nu\beta}=0~; \quad
        \eta_{xb}^{\alpha\beta}=0~; \quad
        \eta_{ab}^{\alpha\beta}=0
                \qquad(a,b\neq x;~\alpha,\beta\neq\theta,\nu)~.
\label{eq:eigenvector7}
\end{eqnarray}
They are visualised in Fig.~\ref{fig:eigenvector7}. The eigenvalue equals
\begin{equation}
        \lambda_7=A_1-2A_2+A_3
\label{eq:eigenwaarde7}
\end{equation}
and has multiplicity $rn(n-3)/2$.

Finally, replica symmetry can also be broken by two spins with different
Roman indices $x$ and $y$. Upon calling the corresponding Greek indices
$\theta$ and $\nu$, this group of eigenvectors reads in its most general
form
\begin{eqnarray}
\fl     \epsilon_x^{\theta\beta}=\epsilon_y^{\nu\beta}=f~; \quad
        \epsilon_x^{\alpha\beta}=\epsilon_y^{\alpha\beta}=g~; \quad
        \epsilon_a^{\alpha\beta}=h~; \quad\nonumber \\
\fl     \eta_{xy}^{\theta\nu}=k~; \quad
        \eta_{xy}^{\theta\beta}=\eta_{xy}^{\alpha\nu}=l~;\quad
        \eta_{xb}^{\theta\beta}=\eta_{ay}^{\alpha\nu}=m~; \quad
        \nonumber \\
\fl     \eta_{xy}^{\alpha\beta}=p~; \quad
        \eta_{xb}^{\alpha\beta}=\eta_{ay}^{\alpha\beta}=q~; \quad
        \eta_{ab}^{\alpha\beta}=t \quad%\nonumber\\
        (a,b\neq x,y;~\alpha,\beta\neq\theta,\nu)~.
\end{eqnarray}
Again (\ref{eq:eigenvalue_equation}) and the orthogonality relations are
used in order to find explicit solutions. One eigenvalue is given by
\begin{equation}
\fl     \lambda_8=(B_1-2B_2+B_3)+2n(B_2-B_3-B_4+B_5)+n^2(B_3-2B_5+B_6)
\label{eq:eigenwaarde8}
\end{equation}
\begin{figure}[t]
%\epsfxsize 8.cm
%\centerline{\rotate[r]{\epsfbox{iegenv8.eps}}}
\centerline{\includegraphics[%height=0.5\textwidth,
        width=8cm,angle=-90]{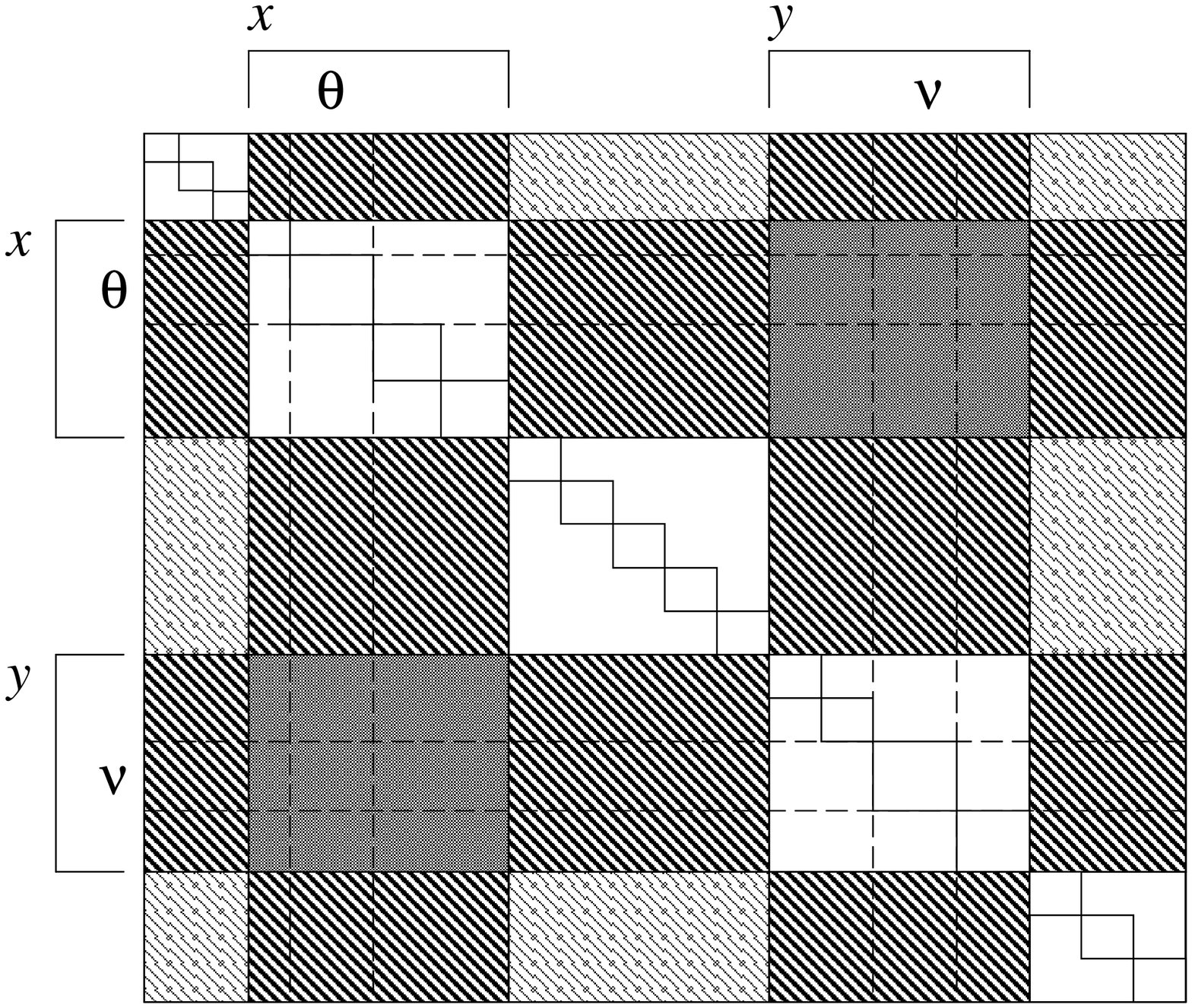}}
\caption{Representation of the eigenvectors (\ref{eq:eigenvector8}).
Empty space denotes zero elements; spaces with the same fill pattern denote
elements with identical matrix elements. Dark spaces denote the components
$\eta_{xy}^{\alpha\beta}$, striped spaces the components
$\eta_{xb}^{\alpha\beta}$ and $\eta_{ay}^{\alpha\beta}$,
the other non-empty spaces the components $\eta_{ab}^{\alpha\beta}$.}
\label{fig:eigenvector8}
\end{figure}
with the corresponding eigenvectors
(Fig.~\ref{fig:eigenvector8})
\begin{eqnarray}
        &&\epsilon_x^{\theta\beta}=\epsilon_y^{\nu\beta}=
        \epsilon_x^{\alpha\beta}=\epsilon_y^{\alpha\beta}=
        \epsilon_a^{\alpha\beta}=0~; \quad
        \eta_{xy}^{\theta\nu}=\eta_{xy}^{\theta\beta}=
        \eta_{xy}^{\alpha\nu}=\eta_{xy}^{\alpha\beta}=k;\quad
        \nonumber \\ &&
        \eta_{xb}^{\theta\beta}=\eta_{ay}^{\alpha\nu}=
        \eta_{xb}^{\alpha\beta}=\eta_{ay}^{\alpha\beta}=
                -\frac1{r-2}~k~; \quad
        \eta_{ab}^{\alpha\beta}=\frac2{(r-2)(r-3)}~k
\label{eq:eigenvector8}
\end{eqnarray}
and with degeneracy $r(r-3)/2$. All $\epsilon$-components vanish.

Yet another eigenvalue is
\begin{equation}
        \lambda_9=B_1-2B_2+B_3
\label{eq:eigenwaarde9}
\end{equation}
\begin{figure}[t]
%\epsfxsize 8.cm
%\centerline{\rotate[r]{\epsfbox{iegenv9.eps}}}
\centerline{\includegraphics[%height=0.5\textwidth,
     width=8cm,angle=-90]{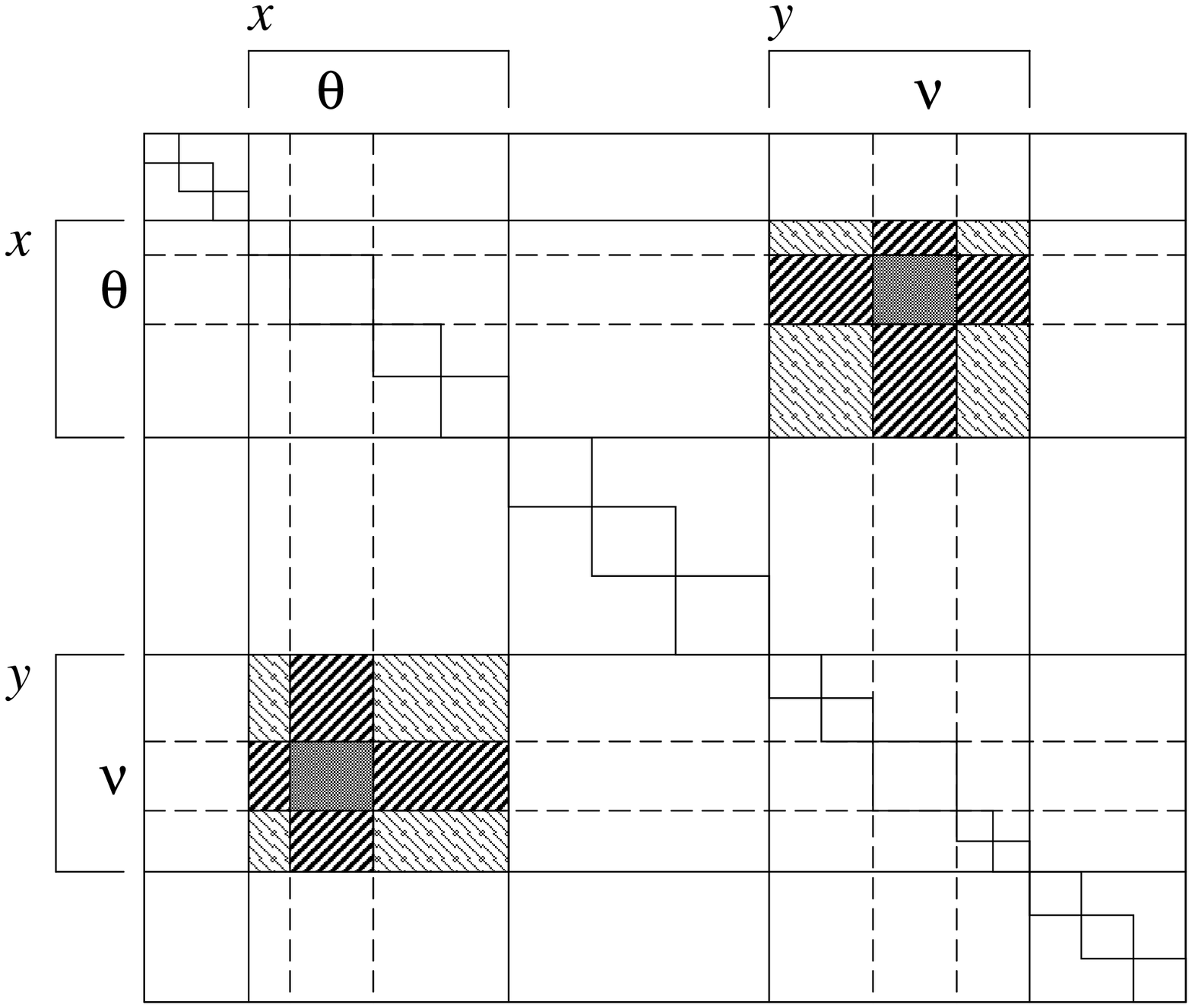}}
\caption{Representation of the eigenvectors (\ref{eq:eigenvector9}).
Empty space denotes zero elements; spaces with the same fill pattern denote
elements with identical matrix elements. The dark spaces denote the
components $\eta_{xy}^{\theta\nu}$, the striped spaces the components
$\eta_{xy}^{\theta\beta}$ and $\eta_{xy}^{\alpha\nu}$, the other non-empty
spaces the components $\eta_{xy}^{\alpha\beta}$.}
\label{fig:eigenvector9}
\end{figure}
with eigenvectors given by
\begin{eqnarray}
        &&\eta_{xy}^{\theta\nu}=k~;\quad
        \eta_{xy}^{\theta\beta}=\eta_{xy}^{\alpha\nu}=-\frac1{n-1}~k~;\quad
        \eta_{xy}^{\alpha\beta}=\frac1{(n-1)^2}~k;\quad
        \nonumber \\ &&
        \epsilon_x^{\theta\beta}=\epsilon_y^{\nu\beta}=
        \epsilon_x^{\alpha\beta}=\epsilon_y^{\alpha\beta}=
        \epsilon_a^{\alpha\beta}=0~; \quad
        \nonumber \\ &&
        \eta_{xb}^{\theta\beta}=\eta_{ay}^{\alpha\nu}=
        \eta_{xb}^{\alpha\beta}=\eta_{ay}^{\alpha\beta}=
        \eta_{ab}^{\alpha\beta}=0\;;
\label{eq:eigenvector9}
\end{eqnarray}
This corresponds to a situation where only the components with the
marked indices $x$ and $y$ are non-vanishing.
The vectors are drawn in Fig.~\ref{fig:eigenvector9}.
The degeneracy of the eigenvalue (\ref{eq:eigenwaarde9}) is $\frac12~r(r-1)(n-1)^2$.

\begin{figure}[t]
%\epsfxsize 8.cm
%\centerline{\rotate[r]{\epsfbox{iegenv10.eps}}}
\centerline{\includegraphics[%height=0.5\textwidth,
    width=8cm,angle=-90]{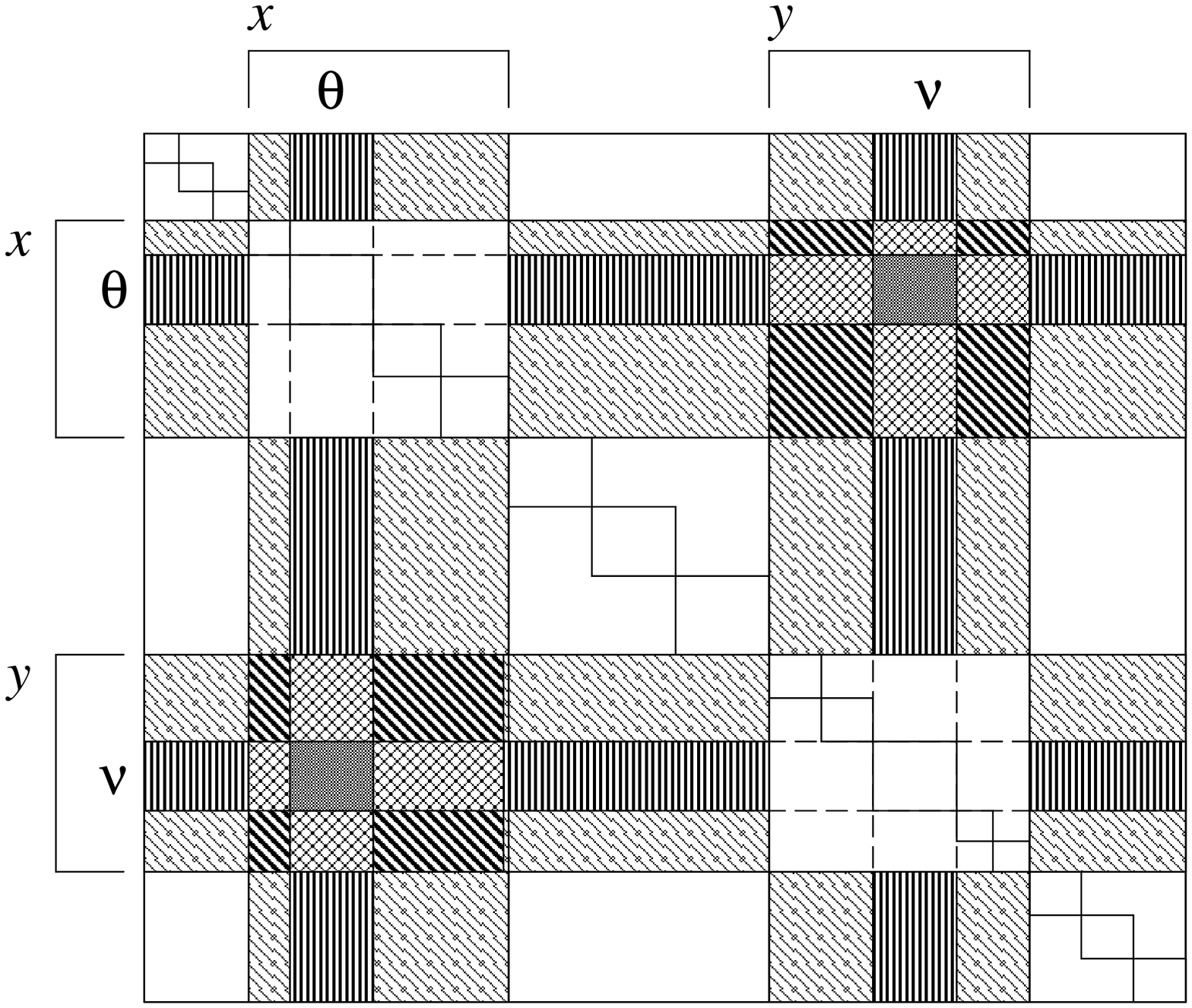}}
\caption{Representation of the eigenvectors (\ref{eq:eigenvector10}).
Empty space denotes zero elements; spaces with the same fill pattern denote
elements with identical matrix elements. The dark spaces denote the
components $\eta_{xy}^{\theta\nu}$, the checked spaces the components
$\eta_{xy}^{\theta\beta}$ and $\eta_{xy}^{\alpha\nu}$,
the diagonally striped spaces the components $\eta_{xy}^{\alpha\beta}$,
the vertically striped spaces the components $\eta_{xb}^{\theta\beta}$ or
$\eta_{ay}^{\alpha\nu}$, the other non-empty spaces the components
$\eta_{xb}^{\alpha\beta}$ or $\eta_{ay}^{\alpha\beta}$.}
\label{fig:eigenvector10}
\end{figure}

Finally, the last eigenvector reads (Fig.~\ref{fig:eigenvector10})
\begin{eqnarray}
        &&\epsilon_x^{\theta\beta}=\epsilon_y^{\nu\beta}=
        \epsilon_x^{\alpha\beta}=\epsilon_y^{\alpha\beta}=
        \epsilon_a^{\alpha\beta}=0~; \quad
        \eta_{ab}^{\alpha\beta}=0\;.
        \nonumber \\
        &&\eta_{xy}^{\theta\nu}=k~;\quad
    \eta_{xy}^{\theta\beta}=\eta_{xy}^{\alpha\nu}=\frac{n-2}{2(n-1)}~k~;
        \quad
     \eta_{xb}^{\theta\beta}=\eta_{ay}^{\alpha\nu}=-\frac1{2(r-2)}~k~;\quad
        \nonumber \\
        &&\eta_{xy}^{\alpha\beta}=-\frac1{(n-1)}~k;\quad
        \eta_{xb}^{\alpha\beta}=\eta_{ay}^{\alpha\beta}=
                \frac1{2(n-1)(r-2)}~k~.
\label{eq:eigenvector10}
\end{eqnarray}
The corresponding eigenvalue is equal to
\begin{equation}
        \lambda_{10}=B_1+(n-2)B_2-(n-1)B_3-nB_4+nB_5
\label{eq:eigenwaarde10}
\end{equation}
and has degeneracy $\frac12r(r-2)2(n-1)$.

\subsection{The multiplicity of the eigenvalues}
\label{app:multiplicity}

This appendix has been included in the present paper since
no information is available in the replica literature
 about explicit methods to find the multiplicity of the
eigenvalues of the Hessian matrix. We do not aim for mathematical
rigour, but just aim to aid the reader by giving a heuristic method for
finding the solution.

\subsubsection{Eigenvalues $\lambda_{3,4}$ and $\lambda_{5,6}$}

We first focus on $\lambda_{5,6}$.
The sum of all $r$ eigenvectors (\ref{eq:eigenvectors5_6}) equals zero,
such that there are at most $r-1$ linearly independent vectors. In the
sequel we show that the rank of the matrix composed of all eigenvectors is
exactly given by this latter value.
We consider the $r$-dimensional sub-matrix constructed by that part
of the $\epsilon$-component of the vectors (\ref{eq:eigenvectors5_6}) with
fixed Greek indices (e.g.\ $\alpha=1,~\beta=2$) viz.\ $(\epsilon_a^{12},~a=1,
\dots,r)$; the $r$ vectors are obtained by varying the Roman index $x$.
This matrix reads
\begin{equation}
  \IM=(f-g)~\II+g~\IP\qquad g=\frac{-1}{r-1}~f
\end{equation}
where $\II$ and $\IP$ are the unit matrix and the projector
matrix, respectively, here both of dimension $r$. Since
these matrices commute they can be
diagonalized simultaneously. The only eigenvector of $\IP$ with a non-zero
eigenvalue is the vector $(1,1,\dots,1)$ with eigenvalue $r$. All other
eigenvectors can be chosen orthogonal to this vector. This results in the
following eigenvalues for $\IM$: one non-degenerated eigenvalue $(f-g)+rg=0$
and an $(r-1)$-fold degenerate eigenvalue $f-g=\frac{r}{r-1}f$.
We conclude that all four eigenvalues (\ref{eq:eigenwaarde5_6})
have degeneracy $(r-1)$.

The multiplicity of $\lambda_{3,4}$ can be found in an analogous way. Since
eigenvectors with a different Roman index $x$ are always
independent, it is sufficient to determine the dimension of the matrix constructed
 from the $n$ eigenvectors with the Roman index fixed, e.g.\ $x=1$,
and with the Greek index $\theta$ running from $1$ to $n$.

\subsubsection{Eigenvalues $\lambda_{7}$ and $\lambda_{8}$}
\label{multiplicity78}

We start with the less complicated calculation for $\lambda_8$. As before we consider
a sub-matrix which is the $r(r-1)/2$-dimensional matrix of the $\eta$-part of the
eigenvectors (\ref{eq:eigenvector8}), with fixed Greek indices
e.g.\ $\alpha=1=\beta$ viz.\
$(\eta_{ab}^{\alpha/\beta},~a<b=1,\dots,r)$
\begin{eqnarray}
  \IM&=&(f-g)~\II+g~\IP+(h-g)~\IB
  \\
  && g=\frac{-1}{r-2}~f\qquad h=\frac{2}{(r-2)(r-3)}~f
  \nonumber \\
  &&\IB_{\alpha\beta,\gamma\delta}=
     (1-\delta_{\alpha\gamma})(1-\delta_{\alpha\delta})
     (1-\delta_{\beta\gamma})(1-\delta_{\beta\delta}),
      ~\alpha<\beta,\gamma<\delta
  \nonumber\,.
\end{eqnarray}
Because the three matrices appearing above commute, the problem of
finding the eigenvalues $\mu_i$ of $\IM$ reduces to finding the eigenvalues
$\gamma_i$ of $\IB$. First, one finds the non-trivial eigenvector of
$\IP$, $(1,1,\dots,1)$, which is also an eigenvector of $\IB$ with
eigenvalue $(r-1)(r-3)/2$. The other eigenvectors $(x_{ab})$, with $a<b$, can be
chosen orthogonal to the trivial one, i.e. $\sum_{a<b}x_{ab}=0$, leading to the
following simplified eigenvalue equation for the eigenvectors of $\IB$
\begin{eqnarray}
  &\gamma x_{ab}=\sum_{c<d}\IB_{ab,cd}~
    x_{cd}=x_a+y_a+x_b+y_b+x_{ab}&
  \nonumber \\
  &\Downarrow& \hspace{-4.cm}x_a=\sum_{b(>a)}x_{ab}
                \quad ~~~~~~ y_b=\sum_{a(<b)}x_{ab}
  \nonumber \\
  &(1-\gamma)x_{ab}=x_a+y_a+x_b+y_b\,.&
  \label{eq:E1}
\end{eqnarray}
The first solution of these equations is $\gamma=1$, with multiplicity $r(r-1)/2-r$ due
to the condition $x_a+y_a=0,~a=1,\dots,r$.
When $\gamma\neq1$, we can sum (\ref{eq:E1}) in two different ways:
\begin{eqnarray}
  \sum_{a(<b)}&:&(1-\gamma)y_b=(b-1)(x_b+y_b)+\sum_{a(<b)}(x_a+x_b)
  \nonumber \\
  \sum_{b(>a)}&:&(1-\gamma)x_a=(n-a)(x_a+y_a)+\sum_{b(>a)}(x_b+y_b)
  \nonumber\,.
\end{eqnarray}
Adding the two equations gives
\begin{equation}
  (3-\gamma-r)(x_a+y_a)=0
  \label{eq:E2}
\end{equation}
and leads to the eigenvalue $\gamma=3-r$. Because of the second factor in
(\ref{eq:E2}) we can conclude that $\IB$ has no other eigenvalues than the
ones already found, leading to the multiplicity $r-1$ for $\gamma=3-r$.
It turns out that only $\gamma=1$ gives a non-zero eigenvalue of $\IM$. We
can therefore conclude that the rank of $\IM$ is equal to
$r(r-1)/2-r=r(r-3)/2$.

The same procedure can be followed to determine the multiplicity of $\lambda_7$,
but now by considering $\IM$ as composed by the $\epsilon$-part of the eigenvectors,
for a fixed choice of the Roman index $x$. Eigenvectors with different
Roman indices are again linearly independent.

\subsubsection{Eigenvalue $\lambda_9$}
\label{multiplicity9}

First, we note that each choice of the Roman indices $x$ and $y$ gives a set of independent
eigenvectors,  so we can limit ourselves to finding the rank of the
matrix generated by the vectors with $x=1=y$
\begin{eqnarray}
  &\eta_{11}^{\theta\nu}=f
  \qquad \eta_{11}^{\theta\beta}=\eta_{11}^{\alpha\nu}=g
  \qquad \eta_{11}^{\gamma\delta}=h&
  \nonumber \\
  &g=\frac{-1}{n-1}~f\qquad h=\frac1{(n-1)^2}~f&
  \nonumber
\end{eqnarray}
which can be written as
\begin{eqnarray}
  &&\IM=(f-g)~\II+g~\IP+(h-g)~\IB
  \nonumber \\
  &&\IB_{\alpha\beta,\gamma\delta}=
  \left\{
  \begin{array}{ll}
  0&\alpha=\gamma
  \\
  \tilde \IB_{\beta\delta}&\alpha\neq\gamma ~{\rm with}~\tilde\IB=\IP-\II
  \end{array}\right.
\end{eqnarray}
where the dimension of $\tilde \IB$ is $n$.
The special structure of $\IB$ allows us to find its eigenvalues
quite easily,
namely: one non-degenerate eigenvalue $(n-1)^2$, one $2(n-1)$-fold
degenerate
eigenvalue $-(n-1)$, and  one $(n-1)^2$-fold degenerate eigenvalue 1.
The first and second of these give a zero eigenvalue for $\IM$, leading to
${\rm rank}~ \IM=(n-1)^2$.

\subsubsection{Eigenvalue $\lambda_{10}$}

Calculating the multiplicity of this eigenvalue involves some more work.
In contrast to the previous sections, we here have to calculate the dimension of
the full matrix of eigenvectors, rather than just the dimension of a suitable sub-matrix.
Upon writing the eigenvectors as
\begin{eqnarray}
\fl     \epsilon_x^{\theta\beta}=\epsilon_y^{\nu\beta}=
        \epsilon_x^{\alpha\beta}=\epsilon_y^{\alpha\beta}=
        \epsilon_a^{\alpha\beta}=0~; \quad
        \eta_{ab}^{\alpha\beta}=0\;.
        \nonumber \\
\fl     \eta_{xy}^{\theta\nu}=f~;\quad
        \eta_{xy}^{\theta\beta}=\eta_{xy}^{\alpha\nu}=g~;
        \quad
        \eta_{xb}^{\theta\beta}=\eta_{ay}^{\alpha\nu}=h~;\quad
        \nonumber \\
\fl     \eta_{xy}^{\alpha\beta}=k;\quad
        \eta_{xb}^{\alpha\beta}=\eta_{ay}^{\alpha\beta}=l;
        \nonumber\\
\fl     g=\frac{n\minus 2}{2(n\minus 1)}~f\quad ~~~~~
        h=-\frac1{2(r\minus 2)}~f\quad ~~~~~
        k=-\frac1{(n\minus 1)}~f\quad ~~~~~
        l=\frac1{2(n\minus 1)(r\minus 2)}~f
        \nonumber
\end{eqnarray}
the matrix of eigenvectors $\IM$ reads
\begin{equation}
\fl  \IM_{ab\alpha\beta,cd\gamma\delta}=\left\{
  \begin{array}{ll}
  f&a=c ~{\rm and}~b=d~{\rm and}~\alpha=\gamma~{\rm and}~\beta=\delta \\
  g&a=c ~{\rm and}~b=d~{\rm and}~(\alpha=\gamma~{\rm or}~\beta=\delta) \\
  k&a=c ~{\rm and}~b=d~{\rm and}~\alpha\neq\gamma~{\rm and}
                               ~\beta\neq\delta \\
  h&{\rm one~subindex~and~the~corresponding~superindex~are~equal}\\
  l&{\rm one~subindex~is~equal~(and~not~the~superindex)}\\
  0&{\rm otherwise}
  \end{array}
  \right.
\end{equation}
It is found to consist of sub-matrices $\IB$ (containing elements
$f,g$ and $k$) on the diagonal, and sub-matrices $\ID_i,~i=1,\dots,4$
(containing elements $h$ and $l$, ordered in four different ways) elsewhere.
All of these sub-matrices have dimension $n^2$.

In order to simplify the problem we first construct the matrix $C$ which is
made up of columns which are orthogonal and normalized eigenvectors of $\IB$. For every
sub-matrix $\ID_i$ we construct $C~\ID_i ~C^T$, where $C^T$ is the transposed matrix
of $C$.  Next we construct the matrix
\begin{equation}
  \hat C=\left(\begin{array}{lcl}
  C&&\\&\ddots&\\&&C
  \end{array}\right)\,.
  \nonumber
\end{equation}
Since $\hat C\hat C^T=\II$ by construction, the eigenvalue equations of $\IM$
and ${\cal M}\equiv\hat C~\IM~\hat C^T$ are the same, and we can restrict
ourselves to solving the simpler eigenvalue problem of the latter matrix.

First we focus on the matrix $\IB$ and the construction of the matrix $C$.
As in Section~\ref{multiplicity9} we have
\begin{eqnarray}
  &&\IB=(f-g)~\II+g~\IP+(k-g)~\IK \nonumber\\
  &&\IK_{\alpha\beta,\gamma,\delta}=
    \left\{\begin{array}{ll}
    0&\alpha=\gamma \\
    \tilde \IK_{\beta\delta}&\alpha\neq\gamma~{\rm with}~\tilde\IK=\IP-\II
    \end{array}\right.
    \nonumber\,.
\end{eqnarray}
Upon using the Gramm-Schmidt procedure for constructing a set of orthogonal and normalized
eigenvectors of $\tilde\IK$, we arrive at the following result:
\begin{equation}
\fl  \tilde C=\left(\begin{array}{ccccccc}
    \frac{1}{\sqrt{n}}   &\frac{1}{\sqrt{n}}   &\frac{1}{\sqrt{n}}
        &\frac{1}{\sqrt{n}}&\frac{1}{\sqrt{n}}&\dots&\frac{1}{\sqrt{n}}\\
    \frac{1}{\sqrt{2}}   &-\frac{1}{\sqrt{2}}  &0                 &0
        &0                 &\dots&0         \\
  \frac12\sqrt{\frac23}&\frac12\sqrt{\frac23}&-\sqrt{\frac23}&0&0&\dots&0\\
    \frac13\sqrt{\frac34}&\frac13\sqrt{\frac34}&\frac13\sqrt{\frac34}&
        -\sqrt{\frac34}&0&\dots&0\\
    \vdots&\vdots&\vdots&\vdots&\vdots&\ddots&\vdots\\
    \frac{1}{n-1}\sqrt{\frac{n-1}{n}}&\frac{1}{n-1}\sqrt{\frac{n-1}{n}}&
    \frac{1}{n-1}\sqrt{\frac{n-1}{n}}&\frac{1}{n-1}\sqrt{\frac{n-1}{n}}&
    \frac{1}{n-1}\sqrt{\frac{n-1}{n}}&\dots&-\sqrt{\frac{n-1}{n}}
  \end{array}\right)\,.
  \label{eq:E4}
\end{equation}
Due to the similar structure of the two matrices $\tilde \IK$ and
$\IK$, we now can immediately read off
the matrix of eigenvectors $C$ of $\IB$:
one takes a matrix with the structure of (\ref{eq:E4}), and multiplies each matrix
element by $\tilde C$, arriving at a matrix with dimension $n^2$.
Given this matrix it is straightforward to calculate $C~\ID_i~C^T$ for all
sub-matrices $\ID_i$. We arrive at
\begin{equation}
\fl  {\cal M}_{ab\alpha\beta,cd\gamma\delta}
   = \left\{\begin{array}{ll}
      \frac{n^2}{2(n-1)}&a=c~{\rm and}~b=d~{\rm and}~
                        \alpha=\gamma=1~{\rm and}~\beta=\delta\neq1\\
                        &a=c~{\rm and}~b=d~{\rm and}~
                        \alpha=\gamma\neq1~{\rm and}~\beta=\delta=1\\
      \frac{-n^2}{2(n-1)(r-2)}&a=c~{\rm and}~b\neq d~{\rm and}~
                        \alpha=\gamma\neq1~{\rm and}~\beta=\delta=1\\
                        &a=d~{\rm and}~b\neq c~{\rm and}~
                        \alpha=\delta\neq1~{\rm and}~\beta=\gamma=1\\
                        &a\neq d~{\rm and}~b=c~{\rm and}~
                        \alpha=\delta=1~{\rm and}~\beta=\gamma\neq1\\
                        &a\neq c~{\rm and}~b=d~{\rm and}~
                        \alpha=\gamma=1~{\rm and}~\beta=\delta\neq1\\
     0                  &{\rm otherwise}
      \end{array}\right.        \,.
\end{equation}
In view of the large number of zero-rows in this matrix, it is convenient to
define a $\frac12r(r-1)2(n-1)$-dimensional matrix $\hat {\cal M}$, which
contains only the non-trivial rows
\begin{eqnarray}
  &&\hat {\cal M}_{ab\alpha,cd\beta}={\cal M}_{ab\gamma\delta,cd\mu\nu}
  \nonumber \\
  &&\begin{array}{lll}
  \gamma=1&\delta=\alpha+1&{\rm for}~\alpha=1,\dots,n-1\\
  \gamma=\alpha-(n-2)&\delta=1&{\rm for}~\alpha=n,\dots,2(n-1)\\
  \mu=1&\nu=\beta+1&{\rm for}~\beta=1,\dots,n-1\\
  \mu=\beta-(n-2)&\nu=1&{\rm for}~\beta=n,\dots,2(n-1)
  \end{array}   \,.
\end{eqnarray}
This matrix can be written as
$\hat {\cal M}=\frac{n^2}{2(n-1)}\II+\frac{-n^2}{2(n-1)(r-2)}{\rm
I\!N}$,
where ${\rm I\!N}$ is a matrix with elements $0$ or $1$ only. As in
Section~\ref{multiplicity78} the eigenvalues of this matrix are obtained
by summing the eigenvalue equation for ${\rm I\!N}$ in two different ways.
We then find the eigenvalue $\lambda=-1$ with multiplicity
$\frac12r(r-2)2(n-1)$,
and the eigenvalue $\lambda=r-2$ with multiplicity $r(n-1)$.
It turns out that only the first of these eigenvalues gives a non-zero eigenvalue for
$\hat{\cal M}$, and we may conclude that ${\rm rank}~\IM=r(r-2)(n-1)$.


\section*{References}

\begin{thebibliography}{99}


\bibitem{CPS} Coolen A C C, Penney R W and Sherrington D 1993
   {\it Phys.~Rev.~B} {\bf 48} 16116

\bibitem{DFM} Dotsenko V, Franz S and M\'ezard M 1994
   {\it J.~Phys.~A: Math.~Gen.} {\bf 27} 2351

\bibitem{PS} Penney R W and Sherrington D 1994
   {\it J.~Phys.~A: Math.~Gen.} {\bf 27} 4027

\bibitem{FD} Feldman D E and Dotsenko V S 1994
   {\it J.~Phys.~A: Math.~Gen.} {\bf 27} 4401

\bibitem{C} Caticha N 1994
   {\it J.~Phys.~A: Math.~Gen.} {\bf 27} 5501

\bibitem{Ho84} Horner H 1984
%   Dynamic mean field theory of the \sk-spin glass,
   {\it Z. Phys. B} {\bf 57} 29

\bibitem{SKb} Sherrington D and Kirkpatrick S 1975
%   Solvable model of a spin glass
   {\it Phys. Rev. Lett.} {\bf 35} 1792

\bibitem{Sh} Shinomoto S 1987
   {\it  J.~Phys.~A: Math.~Gen.} {\bf 20} L1305

\bibitem{PCS} Penney R W, Coolen A C C and Sherrington D 1993
%   Coupled dynamics of fast spins and slow interactions in neural networks
% and spin systems,
   {\it J.~Phys.~A: Math.~Gen.} {\bf 26} 3681

\bibitem{DH} Dong D W and Hopfield J J 1992
   {\it Network} {\bf 3} 267

\bibitem{DF} Dotsenko V S and Feldman D E 1994
   {\it J.~Phys.~A: Math.~Gen.} {\bf 27} L821

\bibitem{Dorotheyev} Dorotheyev E A 1992
   {\it J.~Phys.~A: Math.~Gen.} {\bf 25} 5

\bibitem{LNPS} Lattanzi G, Nardulli G, Pasquariello G and Stramaglia S 1997
   {\it Phys.~Rev.~E} {\bf 56} 4567

\bibitem{CS} Caroppo D and Stramaglia S 1998
   {\it Phys.~Lett.~A} {\bf 246} 55

\bibitem{Kuramoto} Kuramoto Y 1975
   in {\it International symposium on mathematical problems in theoretical
        physics}, H.~Araki, ed. (Springer, New York)

\bibitem{FS94} Fukai T and Shiino M 1994
%   Memory encoding by oscillator death,
   {\it Europhys. Lett.} {\bf 26} 647

\bibitem{AT} de~Almeida J R L and Thouless D J 1978
%  Stability of the Sherrington-Kirkpatrick solution of a spin-glass model,
   {\it J.~Phys.~A: Math.~Gen.} {\bf 11} 983

\bibitem{JBC} Jongen G, Boll\'{e} D and Coolen A C C 1998
%   The \xy~spin glass with slow dynamic couplings,
   {\it J.~Phys.~A: Math.~Gen.} {\bf 31} L737

\bibitem{KS} Kirkpatrick S and Sherrington D 1978
%   Infinite-ranged models of spin-glasses,
   {\it Phys. Rev. B} {\bf 17} 4384

\bibitem{MPV} M\'ezard M, Parisi G and Virasoro M A 1997 {\it Spin Glass
   Theory and Beyond} (World Scientific, Singapore)

\bibitem{Abramowitz} Abramowitz M and Stegun I A (Eds) 1965 {\it
   Handbook of Mathematical Functions} (Dover Publications, New York)

\bibitem{Anemuller} Anem\"uller J 1996
   IPNN MSc Project Report, King's College London

\bibitem{WeS} Whyte W and Sherrington D 1996
%   Replica-symmetry breaking in perceptrons,
   {\it J.~Phys.~A: Math.~Gen.} {\bf 29} 3063

\bibitem{jort} van Mourik J and Coolen A C C 2000 {\em cond-mat/0009151}
(submitted to J. Phys. A)
\end{thebibliography}
\end{document}